\documentclass{aa}  

\usepackage{amsmath}
\usepackage{graphicx}
\usepackage{txfonts}
\usepackage{lipsum}
\usepackage{subcaption}         
\usepackage{lscape}             
\usepackage{placeins}           
                                
\begin{document}

\title{Violent relaxation in hierarchical stellar systems: pathways toward massive black hole formation in the early Universe}

\titlerunning{Violent relaxation in hierarchical stellar systems}

   \author{Dominik R.G. Schleicher\inst{1}\corrauth{dominik.schleicher@uniroma1.it}        
        \and Francesco Flammini Dotti\inst{2,3,1}
        \and
        Mat\'ias Liempi\inst{1}
        \and 
        Tommaso Zana\inst{1}
        \and
        Andr\'es Escala\inst{4}
        \and
        Muhammad A. Latif\inst{5}
        \and
        Roberto Capuzzo-Dolcetta\inst{1}
        \and
        Marcelo C. Vergara\inst{6,4}
        \and
        Lorenzo Paparella\inst{1}
        }

   \institute{Dipartimento di Fisica, Sapienza Universit\`a di Roma, Piazzale Aldo Moro 5, 00185 Rome, Italy
   \and Department of Physics, New York University Abu Dhabi, PO Box 129188 Abu Dhabi, UAE
\and The Center for Astrophysics and Space Science, NYUAD Research Institute, New York University Abu Dhabi, PO Box 129188, Abu Dhabi, UAE
\and Departamento de Astronom\'ia, Universidad de Chile, Casilla 36-D, Santiago, Chile 
\and
Physics Department, College of Science, United Arab Emirates University, PO Box 15551,Al-Ain, United Arab Emirates
\and
Departamento de Astronomía, Facultad Ciencias Físicas y Matemáticas, Universidad de Concepción, Av. Esteban Iturra s/n,
Concepción, Chile
}

   \date{Received September 30, 20XX}

  \abstract
   {Massive stellar systems have been detected in the high-redshift Universe via the James Webb Space Telescope using gravitational lensing techniques. Such stellar systems may evolve on short dynamical timescales, despite a formally large 2-body relaxation time. If the stellar system consists of a large number of gravitationally bound clusters, a 2-body relaxation time can be defined for the star clusters in analogy to the 2-body relaxation time for stars, implying a faster evolution determined by the number of stars per cluster. At their formation, clusters will go through a process of "cold collapse" as part of their early, non-collisional, evolution. This phase enhances the mass segregation between the more massive and less massive population of stars. We derive a criteria based on the mass of the two components that determines under which conditions the massive component may even decouple from the cluster itself. Mergers of clusters and multiple mergers on average tend to produce systems with larger cluster radii. For the formation of massive black holes via dense stellar systems, the most promising pathway consists in the cold collapse of a dense massive cluster, after few mergers (or possibly none). In such a cluster with a sufficient number of massive stars, the massive population of stars will dynamically decouple in the central region and go through core collapse, leading to the formation of a central massive object. In the case of multiple mergers, massive but more spatially extended  stellar systems will form. }


   \maketitle
\nolinenumbers 

\section{Introduction}
Violent relaxation is a rapid process where a collisionless system, such as a newly formed galaxy or star cluster, reaches a stable state of equilibrium. It happens through fast changes in the overall gravitational field, on timescales  corresponding to only  a few crossing times. It is a fundamental process in the evolution of star clusters, which may occur already at formation in case of sub-virial initial conditions (cold collapse) \citep[e.g., ][]{akiyama1989}, as well as during the evolution of star clusters particularly in the case of mergers or in case of strong external gas inflows. 

This makes it particularly relevant as a build-up process of observed massive stellar systems at high redshift. Observational evidence is showing a large number of even massive stellar systems at early times in the Universe.  In particular, JWST has discovered massive star clusters at high redshift, such as the Cosmic Gems at redshift $10.2$ \citep{Adamo2024} or the Firefly Sparkle at redshift $8.3$ \citep{Mowla2024}. More than 220 massive dense clusters were found by JWST via gravitational lensing at redshifts $1\leq z\leq4$ \citep{Claeyssens2026}. These are ideal candidates to form intermediate-mass black holes via stellar collisions \citep[e.g.,][]{Vergara2025a,  Vergara2025b,Rantala2025, Rantala2026,    Schleicher2026, Bocchi2026, Williams2026}. While typical models so far do not explicitly consider the process of violent relaxation, it is likely relevant in setting the initial conditions of star clusters before their subsequent evolution and before potentially going through a core collapse phase \citep[see e.g.,][]{Portegies2002,Katz2015,    Reinoso2018, Reinoso2020, Vergara2021}. Similarly, during the hierarchical  assembly of star clusters in dwarf galaxies and their subsequent mergers, repeated violent relaxation events are expected and will strongly influence the subsequent evolution of these clusters \citep{Lahen2025a, Lahen2025b}.

JWST also discovered a class of objects whose properties are still more uncertainty, the so-called Little Red Dots \citep[e.g.,][]{Matthee2024, Greene2024}. These could be very dense and massive stellar systems \citep[e.g.][]{Guia2024, Escala2025, Liempi2026b}, or Active Galactic Nuclei with a quite massive black hole \citep{Maiolino2024, Maiolino2025, Juod2026}, with typical masses above those expected from the typical black hole mass - bulge mass relation \citep[e.g.][]{Haering2004, Schutte2019}. Even if the real nature and origin of these systems is still unknown, many of the scenarios for their formation involve massive star clusters and efficient stellar collisions \citep[e.g.,][]{Kritos2025, Liempi2026}, implying that violent relaxation could be relevant both for the initial assembly of the stellar masses and to determine the resulting initial conditions. Finally, in the high-redshift Universe, the importance of this process to build up large stellar bulges at high redshift has already been demonstrated via simulations and analytical models by \citet{Bournaud2011} and \citet{Dekel2022}.

For a collisionless system, the evolution of the phase-space distribution function $f(\vec{x},\vec{v},t)$ is given via the collisionless Boltzmann equation,

\begin{equation}
\frac{df}{dt}
=
\frac{\partial f}{\partial t}
+
\dot{\vec{x}}\cdot\nabla_{\vec{x}}f
+
\dot{\vec{v}}\cdot\nabla_{\vec{v}}f=0,\label{Boltzmann}
\end{equation}
where $\vec{x}$ is the coordinate in physical space, $\vec{v}$ is the velocity in the phase-space, and $\dot{\vec{v}}=-\nabla\Phi(\vec{x})$, where $\Phi(\vec{x})$ is the gravitational potential. According to Liouville's theorem, the phase-space volume will be conserved during collisionless evolution, while for a collisional system, we no longer have $df/dt=0$ along trajectories. Effectively, this theorem holds for the fine-grained phase-space density, which is conserved along trajectories. To describe the evolution of the macrostate, \citet{LyndenBell1967} has introduced a coarse-grain description considering the number of stars within larger cells in the phase-space volume. The process of violent relaxation involves a strong redistribution of the matter. As a result, the gravitational potential will change rapidly, and the energy per unit mass of the stars, $\epsilon$, will change rapidly as\begin{equation}
\frac{d\epsilon}{dt}=\frac{\partial \Phi}{\partial t},
\end{equation}
driving the phase-space mixing processes. While the framework of \citet{LyndenBell1967} considered 
an extended framework of initial conditions, some of the main results can be understood even assuming a so-called "water-bag" initial condition, with a fixed phase-space density $\eta_0$ in a spherical phase-space volume. As a result, during collision-less evolution, each microstate can be either equal to $\eta_0$ or $0$. The phase space number density of particles cannot grow larger than $\eta_0$, presenting a classical analog to the Pauli exclusion principle. Considering macrocells $i$ consisting of a number $\nu$ of microcells, the average phase-space distribution function of the macrocell $\bar{f}_i=n_i \eta_0/\nu$, where $n_i$ is the number of occupied microcells in the macrocell. After efficient mixing, so considering a maximum entropy state as well as mass and energy conservation, one can show that the mean distribution function corresponds to the Fermi-Dirac distribution with $
\bar{f}=\frac{\eta_0}{1+\mathrm{exp}(\beta(\epsilon-\mu))}$,
where $\beta=1/(k_BT)$ is the inverse temperature of the stellar system, with temperature $T=\frac{m\sigma^2}{k_B}$, $\sigma$ the velocity dispersion, $m$ the mass per star and $k_B$ the Stefan-Boltzmann constant and $\mu$ the chemical potential per unit mass. In the dilute limit, so for $\bar{f}\ll 1$, one obtains the Maxwell-Boltzmann approximation
$
\bar{f}=\eta_0 \mathrm{exp}(-\beta(\epsilon-\mu))$.

Importantly, we have always $\bar{f}\leq \eta_0$, so the initial phase space density is providing an upper limit on the coarse-grain phase space density that can be reached during the dynamical evolution. \citet{Tremaine1986} provided a description of the mixing processes during violent relaxation, showing that elliptical galaxies could form from a cold or clumpy initial state through collisionless evolution.  A first application of this scenario comes from \citet{akiyama1989}. In one of the first direct N-body simulations of rotating star clusters, they found that a same-mass stellar system in a sub-virial state goes through a fast violent relaxation phase. It brings the system quickly in virial equilibrium within approximately two crossing times. \citet{Farouki1983} pursued N-body experiments to determine mass segregation during violent relaxation, finding some enhancement compared to virialized systems. However, they also found this enhancement to decrease with the number of particles, suggesting it to be driven by 2-body encounters. \citet{flamminidotti2026} instead simulate two separate populations, where $M_1 \gg M_2$, where $M_1$ and $M_2$ are the two total masses of the populations. If the second population has a sub-virial state (while the first is in virial equilibrium), we experience violent relaxation similarly to the case of \citet{akiyama1989}, where the second population relaxes in $\approx$ 2 crossing times.
For a state-of-the-art review on violent  theory, including criticisms and possible extensions, we refer to \citet{Bindoni2008}. We also refer to \citet{Kuli1997} for a slightly reformulated framework, improving on the treatment of the coarse-grain phase space and avoiding an explicit dependence on the stellar masses in the resulting equilibrium configurations. 

\begin{figure}[!h]
    \centering
\includegraphics[scale=0.5]{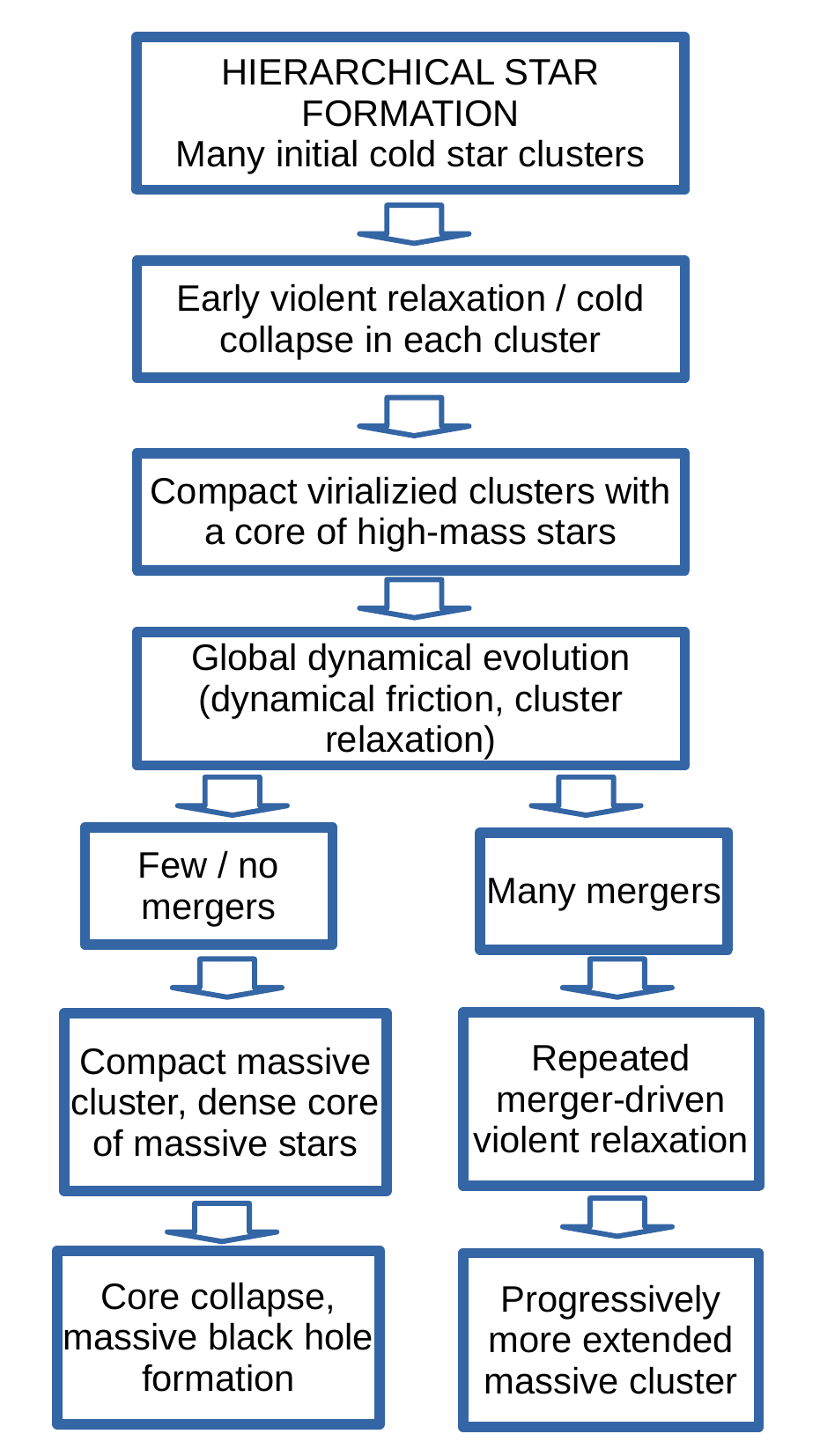}
    \caption{Sketch of the dynamical evolution of a hierarchical stellar system, considering the formation of initially cold star clusters, violent relaxation due to cold collapse in each of the individual clusters, potentially leading to the formation of a compact core of massive stars. The global evolution of the system of star clusters happens due to dynamical friction and relaxation. The subsequent evolution in the case of only few or no mergers leads to the formation of a compact massive cluster, potentially well-suited to form a massive black hole via core collapse. Alternatively, in case of multiple mergers, the build-up of a massive and extended stellar system can occur. An intermediate-mass black hole may still be present depending on the previous evolution of the clusters participating in the merger sequence.}
    \label{fig:violentrelaxation}
\end{figure}

In this paper, our goal is to analyse the role of violent relaxation during the build-up of massive star clusters in the early Universe, including the potential to form massive black holes via these clusters. For this purpose, we start in section~\ref{collisional} with considerations of the global evolution of a hierarchical system of star clusters. In section~\ref{coldsimple}, we provide a reference case for a cold collapse of a star cluster consisting of one single stellar population. In section~\ref{coldcollapsemassive}, we then analyse violent relaxation processes in clusters with normal and massive stars, considering both cold collapse and mergers between clusters. Particularly for the case of cold collapse, we show that violent relaxation may enhance an initially present mass segregation in these clusters, leading to the formation of a compact and dynamically decoupled core of massive stars. In section~\ref{massivecluster}, we consider the build-up of a massive cluster through multiple violent relaxation events, showing how repeated merger processes lead to the formation of more extended configurations, while histories involving few or no mergers might be particularly favourable to build up compact systems that could subsequently form a massive black hole via runaway collisions. A final discussion and conclusion is provided in section~\ref{conclusions}. The general evolutionary sequence we consider in this paper is outlined in Fig.~\ref{fig:violentrelaxation}.

\section{Collisional processes in systems with many star clusters}\label{collisional}

Stellar systems that consist of a large number of stars are often assumed to evolve over very long timescales. The reason is the 2-body relaxation timescale, which is given as \citep{Spitzer1987}: \begin{equation}
t_{\rm relax}=\frac{0.138 N}{\ln \left(\Lambda N\right)}t_{\rm cross},\label{relax}
\end{equation}
where $N$ is the number of stars, $t_{\rm cross}=R/\sigma$ is the crossing time, $R$ the size of the stellar system, $\sigma$ is its velocity dispersion and $\ln\left(\Lambda N\right)$ is the Coulomb logarithm with $\Lambda$ being the Coulomb factor. The latter strongly changes depending on the mass distribution in the cluster. We provide a summarized version of this dependence  in Appendix~\ref{app:columb}. For large stellar systems, this timescale is already very long. For example, \citet{Claeyssens2026} reported more than 220 massive star clusters detected at high redshift with JWST, with masses of $\sim10^5-10^8$~M$_\odot$, and half-mass radii of $\sim1-20$~pc. Taking a characteristic example with mass $10^7$~M$_\odot$ and assuming a mass per star of $1$~M$_\odot$, we have $N=10^7$. Adopting a typical radius of $R=5$~pc, we can estimate the velocity dispersion from virial equilibrium as $\sigma\sim\sqrt{GM/R}\sim94$~km/s. This implies $t_{\rm cross}\sim5.1\times10^4$~years and $t_{\rm relax}\sim3.1\times10^9$~years. This is considerably larger than the Hubble time at redshift 4 and even still comparable to the Hubble time at redshift 1, so one might conclude that it would be difficult for such clusters to evolve via  stellar-dynamical processes. It is now important to consider that even small galaxies like the Cosmic Gems \citep{Adamo2024} or the Firefly Sparkle \citep{Mowla2024} consist of at least 5-10 massive star clusters with typical separations of $20-40$~pc. Cosmological zoom-in hydrodynamics simulations by \citet{Liempi2026LRD} have shown that a few times $10^8$~M$_\odot$ of stars can accumulate on scales of less than $100$~pc in the center of high-redshift galaxies, particularly in the regime when the star formation efficiency is high and feedback relatively inefficienty \citep[see also][]{Dekel2025}. A naive estimate as above then implies a velocity dispersion of $115$~km/s, a crossing time of $8.3\times10^5$~years and a relaxation time of $1.3\times10^{12}$~years, implying evolution times longer than the age of the Universe. We may now consider that the timescale for mass segregation is given as \citep{Binney2009}\begin{equation}
t_{ms}=\frac{\langle m\rangle}{m_{\rm max}}t_{\rm relax},\label{ms}
\end{equation}
where $\langle m\rangle$ is the average mass per star and $m_{\rm max}$ the maximum mass per star. Depending on the stellar initial mass function, the mass segregation time could thus be smaller by a factor of $10-100$, but even in this case would be still at least comparable to the Hubble time at a high redshift.

However, in the discussion so far we have neglected the presence of sub-structure in stellar systems. Stars are known to form in clusters, and we adopt a typical power-law distribution for the cluster mass function as \begin{equation}
\frac{dN}{dm}=Cm^{-\alpha},
\end{equation}
with $\alpha\sim2$ and $C$ a normalization constant that is determined by the total stellar mass in clusters:\begin{equation}
\epsilon_c M_{\rm *,tot}=\int_{m_{\rm cmin}}^{m_{\rm cmax}}\frac{dN}{dm}mdm =\int_{m_{\rm cmin}}^{m_{\rm cmax}}Cm^{-\alpha+1}\,dm =\begin{cases}C\ln\left(\frac{m_{\rm cmax}}{m_{\rm cmin}}\right) , \, \alpha=2\\
\frac{C}{-\alpha+2}\frac{m_{\rm cmax}^{-\alpha+2}}{m_{\rm cmin}^{-\alpha+2}}.\,\alpha\neq2.
\end{cases}
\end{equation}
where $M_{\rm *,tot}$ is the total stellar mass and $\epsilon_c$ the fraction of stars in clusters, $m_{\rm cmin}$ the minimum cluster mass, and $m_{\rm cmax}$ the maximum cluster mass. For $\alpha=2$, the number of clusters is then given as\begin{eqnarray}
N_c=\int_{m_{\rm cmin}}^{\rm m_{\rm cmax}} \frac{dN}{dm}dm&=&\int_{m_{\rm cmin}}^{\rm m_{\rm cmax}} \frac{\epsilon_c M_{\rm *,tot}}{\ln(m_{\rm cmax}/m_{\rm cmin})}m^{-2}dm\\
&=&\frac{\epsilon_c M_{\rm *,tot}}{\ln(m_{\rm cmax}/m_{\rm cmin})} \left( m_{\rm cmin}^{-1}-m_{\rm cmax}^{-1} \right)
\end{eqnarray}
and the average star cluster mass is\begin{equation}
m_{\rm cav}=\frac{\epsilon_c M_{\rm *,tot}}{N_c}=\frac{\ln(m_{\rm cmax}/m_{\rm cmin}) }{m_{\rm cmin}^{-1}-m_{\rm cmax}^{-1}}.
\end{equation}
Very similar expressions can be obtained for $\alpha\neq2$, but we are here only interested in deriving an order of magnitude for the average star cluster mass. Adopting $\alpha=2$, $m_{\rm cmax}=10^7$~M$_\odot$ and $m_{\rm cmin}=10^3$~M$_\odot$, we have $m_{\rm cav}\sim9.2\times10^3$~M$_\odot$. 
As star clusters can spiral to the center of a stellar system via dynamical friction in the same way as massive stars, we can evaluate Eq.~\ref{ms} using the mean and maximum stellar mass in the star clusters, providing a suppression factor of order $10^{-3}$.
We note that star clusters have normally a broken power law, like \citet{Kroupa2001}. The broken power law predicts a sharp increase in small, low-mass objects compared to massive ones. Conversely, the unbroken power law uses a single, fixed rate that results in a more evenly spread, flatter distribution.
This is however not yet the full picture, as Eq.~\ref{ms} is still based on the 2-body relaxation time given via Eq.~\ref{relax}. That expression for the relaxation time is only valid for stellar systems with a homogeneous density distribution. As we discussed, a large fraction of stars in the system forms part of star clusters, corresponding to a highly inhomogeneous distribution where this assumption breaks down. On the other hand, the motion of each star cluster approximately can be regarded as a point particle. If we assume that the star clusters are homogeneously distributed in the stellar system, a relaxation time can be derived for the system of star clusters in an analogous way to the relaxation time for stars, leading to an expression highly similar to Eq.~\ref{relax}, but replacing the number of stars with the number of star clusters as\begin{equation}
t_{\rm relax,c}=\frac{0.138N_c}{\ln (\Lambda N_c)}t_{\rm cross}.\label{relaxc}
\end{equation}
\begin{figure}[!h]
    \centering
\includegraphics{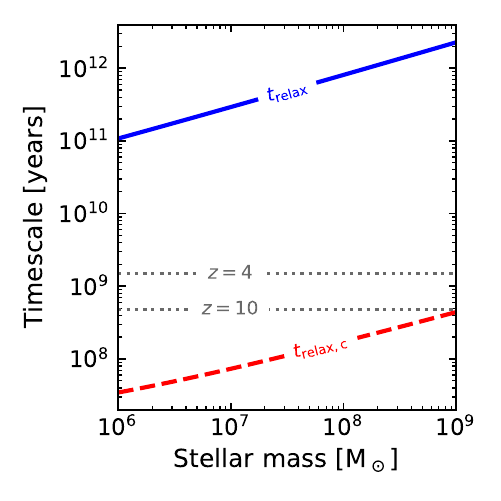}
    \caption{ Solid  blue line shows the relaxation time for single star clusters (Eq. \ref{relax}) assuming a size of $100$\,pc and average stellar mass of $m_\star=1$\,M$_\odot$. On the other hand, as a dashed red line we show the relaxation time of a system of stellar clusters (Eq. \ref{relaxc}) assuming again a size of $100$\,pc and average mass of clusters $m_{\rm cav}=9.2\times 10^3$\,M$_\odot$. For comparison, we include the  age of the Universe at redshift 4 and 10 as dotted gray lines.}
    \label{fig:relax}
\end{figure}

A comparison of this timescale with the standard 2-body relaxation time is provided in Fig.~\ref{fig:relax}. In the system considered here, we have $\sim10^3$ star clusters, thus yielding a star cluster relaxation time of $\sim1.2\times10^7$~years. This timescale is thus considerably shorter than the Hubble time and suggests a rapid evolution of the stellar system. In particular, we may expect that other well-known results for star clusters could at least approximately hold for a cluster of star clusters, as long as each individual cluster is small compared to the size of the system. The global stellar system may then for example go through core collapse, where the timescale in case of equal mass clusters would be $\sim15 t_{\rm relax,c}$ \citep{Cohn1980}. For real systems with a cluster mass distribution, as discussed above, the in-spiral of the most massive clusters might proceed even faster on the mass segregation time \begin{equation}
t_{\rm ms,c}=\frac{m_{\rm cav}}{m_{\rm cmax}}t_{\rm relax,c},
\end{equation}
at least if the number of clusters is large enough for the dynamical friction time to present a good enough approximation. Of course the approximation here should be expected to break down when two star clusters come sufficiently close so that their separation will be comparable to the size of the individual clusters. At that point, however, they will experience the effect of a rapidly changing gravitational potential due to the inspiral of the other clusters, giving rise to violent relaxation. 

We also note that we have assumed here that the number of star clusters is sufficiently large so that their evolution can be described as a gradual evolution due to weak gravitational scattering. This assumption is justified when the number of star clusters is sufficiently high. If the number of clusters is only small or moderate, though, the evolution can be even faster, as the system may then globally go through a violent relaxation event, and evolve on only a few crossing timescales. We provide a criterium for this transition in appendix~\ref{app:global_violent_relaxation} which describes how it depends on the number of clusters. 

\section{Cold collapse in simple star clusters}\label{coldsimple}

We consider simple star clusters with sub-virial conditions, leading to cold collapse. It is perhaps the most simple example of a violent relaxation process showing how a considerable contraction could be achieved over a short timescale.

We consider a Plummer sphere consisting of equal mass stars with mass $m$, defined via the following distribution function \citep{PLUMMER1911, Binney2009} \begin{equation}
f(\vec{x},\vec{v})=\frac{24\sqrt{2}}{7\pi^3}\frac{a^2}{G^5M^4}\left(-E(\vec{x},\vec{v})  \right)^{7/2},\label{fPlummer}
\end{equation}
with $a$ the Plummer radius and the specific energy\begin{equation}
\epsilon(\vec{x},\vec{v})=\frac{1}{2}\vec{v}^2+\Phi_{\rm P}(\vec{x}).
\end{equation}
$\Phi_P$ is the Plummer potential given as\begin{equation}
\Phi_{\rm P}(\vec{x})=-\frac{GM}{\sqrt{r^2+a^2}}.
\end{equation}
The density profile is then\begin{equation}
\rho(r)=\frac{3M}{4\pi a^3}\left(1+\frac{r^2}{a^2}  \right)^{-5/2}.
\end{equation}
The Plummer radius $a$ is related to the half-mass radius as $a=0.766R_h$. In virial equilibrium, the one-dimensional velocity dispersion would be \begin{equation}
\sigma_{\rm Pl}^2(r)
=
\frac{GM}{6\sqrt{r^2+a^2}}.\label{sigma}
\end{equation}
We define the virial ratio as $q=2K/|W|$. Assuming a cold initial condition, we rescale with the virial parameter as\begin{equation}
\sigma^2(r)
=
q\,\sigma_{\rm Pl}^2(r).
\end{equation}Assuming a system with an initial phase space density $\eta_0$, the maximum number density at a given time can be estimated in the \citet{LyndenBell1967} framework via the velocity dispersion $\sigma$ at that time. For a system with radius $R$ and velocity dispersion $\sigma$, a simple estimate of the phase-space volume is thus $R^3\sigma^3$. Considering the constant phase space density $\eta_0$, we then have $n_{\rm max}\sim \eta_0\sigma^3$. Assuming further that the initial number density was $n_i$ and the initial velocity dispersion $\sigma_i$, we have \begin{equation}
n_{\rm max}\sim n_i\left( \frac{\sigma}{\sigma_i} \right)^3.\label{ndens}
\end{equation}
To estimate the gravitational potential and the velocity dispersion in the central region of the Plummer sphere, we consider
\begin{eqnarray}
\Phi(0)&\sim& -\frac{GM}{a},\\
\sigma^2(0)&\sim& \frac{qGM}{6a}.
\end{eqnarray}
Taking $\vec{v}^2\sim3\sigma^2(0)$ and inserting both expressions in Eq.~\ref{fPlummer} and dividing by the mass per star $m$ to normalize as a characteristic phase-space number density, we obtain
\begin{equation}
\eta_0=\frac{24\sqrt{2}}{7\pi^3 m}\sqrt{\frac{(1-\frac{q}{4})^7}{a^3G^3M}}.
\end{equation}
For a Plummer sphere with virial parameter $q$, the total energy integrated over the Plummer sphere can be written as\begin{equation}
E_{\rm tot}=-(1-\frac{1}{2}q)\frac{3\pi GM^2}{32a}.\label{Etot}
\end{equation}
For the initial distribution, we consider a Plummer sphere with $q_{\rm i}$ and $a_{i}$, and we will assume that violent relaxation produces at least approximately a Plummer-distribution with a new Plummer radius $a_{\rm f}$ in a virialized state with $q_{f}=1.$ Energy conservation then implies\begin{equation}
a_f=\frac{a_i}{2-q},\label{afcold}
\end{equation}
implying a ratio of $\sim1/2$ for cold initial conditions. In appendix~\ref{King}, we show how this result generalizes if instead a King or Dehnen model is considered. Using Eq.~\ref{sigma}, the velocity dispersion in the central region of the final Plummer sphere is given as\begin{equation}
\sigma_f^2=\frac{(2-q_i)GM}{6a_i}=\frac{2-q_i}{q_i}\sigma_i^2\label{sigmacold}.
\end{equation}
For $q_i\ll1$, the squared velocity dispersion thus increases by a factor $\sim2/q_i$. From Eq.~\ref{ndens}, the maximum density will then increase by a factor $(2/q_i)^{3/2}$.

\section{Violent relaxation in the presence of massive stars}\label{coldcollapsemassive}

Realistic star clusters have different stellar populations, including many stars with about the average mass in the stellar system but also, at least in terms of the mass fraction, a relevant component of quite massive stars. In the context of standard two-body relaxation theory, it is well known that a population of massive stars can go through a Spitzer instability and lead to a decoulping of the massive component \citep{Spitzer1969}. In this section, we will show that a very similar phenomenon could occur in the context of incomplete violent relaxation. For this purpose, we will consider one limiting case of complete mixing, where both components are very efficiently mixed, as well as the more realistic possibility that mixing is likely going to be incomplete due to the short times during which the gravitational potential changes significantly, and as a simplifying assumption, we assume that the energy of each of the two components is at least approximately conserved. As a result, we will show that the violent relaxation process could potentially give rise to a spatial decoupling of the two mass components, producing a situation where the massive stars are strongly concentrated in the center. 

\subsection{Basic parameters and motivation}

We consider a stellar system with at least two components of stars. The first component consists of the more frequent stars with a mass $m_1$, and a second  component with a larger mass $m_2\gg m_1$. Each component has a respective phase space distribution function $f_1$ and $f_2$ and obeys the collisionless Boltzmann equation as defined in Eq.~\ref{Boltzmann}.

We note that both components experience the same total gravitational potential $\Phi$, which is determined via the Poisson equation:
\begin{equation}
\nabla^2\Phi
=
4\pi G(\rho_1+\rho_2),
\end{equation}
where $\rho_1=\int f_1 m_1 d^3v$ and $\rho_2=\int f_2 m_2 d^3v$ are the respective mass densities of the two components. Considering Liouville's theorem, the phase-space volume of both components will be conserved if initially the two stellar components occupy the same phase space and are described by identical normalized  phase-space distribution functions. 

There are many indications for initial mass segregation already as a result of the star formation process. For example, \citet{Bonnell1998}  argue that the massive stars in the Orion Nebula Cluster are too young to have reached the center through two-body relaxation, implying that their central position originated as a result of the star formation process. The distribution of massive stars in the Orion Nebula Cluster was also independently looked at by \citet{Hillenbrand1998}, showing strong indications of mass segregation. \citet{Pang2013} have demonstrated the presence of mass segregation in HD 97950 and found slightly lower velocity dispersion for higher-mass stars, suggesting them to be dynamically colder. 

These correlations are not only found observationally, but are also theoretically understood. Numerical simulations show that massive stars form preferentially at the centers of star-forming clusters, as gas is funneled towards the center of the clusters via the gravitational potential \citep{Bonnell2003, Smith2009, Peters2010}. In this case, it is very naturally expected for massive stars to be dynamically colder than the typical ones. Taken together, this makes it quite plausible that the massive component could occupy a smaller volume in phase space compared to the standard component, and in this case, we could expect violent relaxation to further enhance the original difference as both components will be able to achieve different final configurations.

The total mass in the less massive population of stars is $M_1=N_1m_1$, with $N_1$ the number of normal stars, and the total mass in massive stars is $M_2=N_2m_2$, with $N_2$ the number of massive stars. We consider the possibility of an initial mass segregation, so we have two different half-mass radii $R_i$ for each component , with $R_2<R_1$. The virial parameters $q_i$ of both components might be different, but in both cases we assume $q_i\ll1$, correspondiing to cold initial conditions.

\subsection{Evolution of mass segregation during cold collapse}
We will analyse the expected results of violent relaxation in the limiting cases of complete mixing (enough time for violent relaxation to be completed) versus incomplete mixing, assuming inefficient energy exchange between the two components.

\subsubsection{Complete mixing}

We assume here that each of the two components is initially described as a subvirial Plummer sphere with the respective Plummer radii $a_1$ and $a_2$ and virial parameters $q_1$ and $q_2$. After complete mixing, both components will follow the same distribution, so they will both correspond to a virialized Plummer sphere with one single Plummer radius $a_f$. To obtain more definite expressions, we neglect here the gravitational interaction energy between the two Plummer spheres, which essentially will provide a negative contribution and deepen the resulting potential. As the gravitational interaction energy is proportional to the product of the two masses, its contribution is relevant and comparable to the self-energy for $M_1\sim M_2$, while it is a minor correction if $M_2\ll M_1$. We note that the effects of the interaction term will be analyzed in greater detail in subsection~\ref{starclustermergers}.

In terms of energy conservation, this means that\begin{equation}
-\left(1-\frac{q_1}{2}\right)\frac{3\pi GM_1^2}{32a_1}-\left(1-\frac{q_2}{2}\right)\frac{3\pi GM_2^2}{32a_2}=-\frac{3\pi G(M_1+M_2)^2}{64a_f}.
 \end{equation}
Thus we obtain as the radius for the mixed Plummer sphere\begin{equation}
a_f=\frac{1}{2}\frac{(M_1+M_2)^2a_1a_2}{(1-\frac{q_1}{2})M_1^2a_2+(1-\frac{q_2}{2})M_2^2a_1},\label{finalradiuscompletemixing}
\end{equation}
In the limiting case where $q_1\sim q_2\sim q$ and $M_2=\delta_M M_1$, we have
\begin{equation}
a_f=\frac{1}{2}\frac{(1+\delta_M)^2a_1a_2}{(1-\frac{q}{2})(a_2+\delta_M^2 a_1)}\sim \frac{1}{2}\frac{(1+\delta_M)^2a_1a_2}{a_2+\delta_M^2 a_1},
\end{equation}
where the second expression corresponds to the limiting case with $q\ll 1$. Let us initially consider the limiting case $\delta_M\ll1$, implying $M_2\ll M_1$. In this case by linearization one can show\begin{equation}
a_f\sim\frac{1}{2}(1+2\delta_M) a_1,
\end{equation}
where all terms quadratic in $\delta_M$ have been neglected. In this case complete mixing will imply that the Plummer radius roughly decreases by a factor of 2, as in the limiting case of a cold single cluster.
On the other hand, if we adopt the limiting case where $\delta_M\sim1$ and $a_1=a_2\sim a_{i}$, we find\begin{equation}
a_f=\frac{1}{2}\frac{(1+\delta_M)^2}{1+\delta_M^2}a_{i},
\end{equation}
where for $\delta_M=0.5$, it implies $a_f/a_{i}\sim0.9$, allowing only for a very moderate reduction of the Plummer radius. We stress that this result is obtained neglecting the gravitational interaction energy between the two initial Plummer spheres.

\subsubsection{Incomplete mixing}

While we have seen that in the case of complete mixing, effects of mass segregation would be destroyed, it has been frequently noted in the literature that mixing will likely not be complete and violent relaxation will not lead to a maximum entropy state, particularly given that the process only proceeds over a few crossing times. To account for that, we here consider the possibility of rather incomplete mixing and approximate conservation of the energy of the massive and the lower-mass component. In that case, we would have two initial Lagrange multipliers $\beta_{1i}\neq\beta_{2i}$, and the average phase-space functions in the dilute regime could be written as
\begin{equation}
\bar{f}_{ki}=A_{ki}\mathrm{exp}(-\beta_{ki}\epsilon),\qquad k=1,2,
\end{equation}
with $A_{ki}$ being the normalization factors and $\beta_{ki}=1/\sigma_{ki}^2$ is the inverse temperatures  following Eq.~\ref{sigmacold}. As discussed above, we assume here that the energies of the two different components are separately conserved. Particularly, if the massive component starts out as colder and somewhat more centrally concentrated, it might collapse earlier than the low-mass component and within a somewhat different volume. Also, as violent relaxation is a fast process occuring within only few crossing times, very little energy could be exchanged via two-body relaxation processes.

Assuming for simplicitly that each mass distribution will form an approximately virialized Plummer distribution, then for each of the Plummer spheres the respective ratios between the initial and final Plummer radii will be given via Eq.~\ref{afcold} and the ratio of their velocity dispersions via Eq.~\ref{sigmacold}. We briefly recall that the half-mass radii are given as $1.305$ times the Plummer radii. The central densities of the two components are
\begin{eqnarray}
\rho_1&=&\frac{M_1}{4\pi a_1^3/3},\\
\rho_2&=&\frac{ M_2}{4\pi a_2^3/3}.
\end{eqnarray}

The central region becomes dominated by the heavy component if $\rho_2>\rho_1$, in other words if  $M_2/a_2^3>M_1/a_1^3$.

In the central region, where the density of both components is approximately constant, the potential should be approximately harmonic,
\begin{equation}
\Phi(r)-\Phi(0)
\simeq
\frac12\Omega^2 r^2
\end{equation}
with an appropriate constant $\Omega$.
To understand the implications for the spatial distribution of the two components, in the final state we assume a Boltzmann distribution in phase space,
\begin{equation}
\bar{f}_{kf}=A_{kf}\mathrm{exp}(-\beta_{kf}\epsilon),
\end{equation}
with \begin{equation}
\beta_{kf}=\frac{1}{\sigma_{kf}^2}=\frac{q_k}{(2-q_k)\sigma_k^2}\label{beta}
\end{equation}
 the inverse temperature of the two components, $\sigma_{kf}$ the velocity dispersion of the two components and $A_{kf}$ an appropriate normalization factor for the two distributions.

Integrating over velocity space, it implies
\begin{equation}
\rho_i(r)
=
\rho_i(0)
\exp\left[
-\frac{\Omega^2 r^2}{2\sigma_{if}^2}
\right].\label{expdens}
\end{equation}
Within the central region, the density profiles should thus be given approximately by Eq.~\ref{expdens} and the description via Plummer spheres should break down on small enough scales. We can  define  characteristic radii for each component as $r_j=\sigma_{jf}/\Omega$ with $r_1/r_2=\sigma_{1f}/\sigma_{2f}$. Then, the condition $\rho_2>\rho_1$ can be rephrased as \begin{equation}
M_2\sigma_{2f}^{-3} > M_1\sigma_{1f}^{-3}\label{crit}
\end{equation}
or with Eq.~\ref{beta}:
\begin{equation}
M_2\sigma_{2i}^{-3} \left( \frac{2-q_{2i}}{q_{2i}} \right)^{-3/2} > M_1\sigma_{1i}^{-3}\left( \frac{2-q_{1i}}{q_{1i}} \right)^{-3/2}.\label{critini}
\end{equation}
While this condition should be evaluated for the specific parameters in an environment, it shows in particular that the probability of decoupling can be enhanced in cases where the massive component is much colder than the normal one, that is when $q_{2i}\ll q_{1i}$.

As the condition for such a decoupling depends strongly on the masses $M_1$ and $M_2$ and particularly the ratio $M_1/M_2$, we aim to get a rough estimate on this ratio. For this purpose, we assume a Salpeter-type initial mass function (IMF) \citep{Salpeter1955} with 
$\frac{dN}{dM}\propto M^{-2.35}$. The number of stars $N(>m_i)$ above a mass threshold $m_i$ then scales as \begin{equation}
N(>m_i)\propto\int_{m_i}^{m_{\rm max}}M^{-2.35}dM\propto m_i^{-1.35}
\end{equation}
for $m_i\ll m_{\rm max}$. Thus, for example in case of $m_1=1$~M$_\odot$ and $m_2=100$~M$_\odot$, we have a ratio $N(>100\, {\rm M}_\odot)/N(>1\,{\rm M}_\odot)\sim100^{-1.35}\sim2.0\times10^{-3}$. While from the point of view of number of stars, the massive component may thus seem negligible, we must also consider the stellar mass $M(>m_i)$ above threshold $m_i$. In this case, we obtain\begin{equation}
M(>m_i)\propto \int_{m_i}^{m_{\rm max}}M^{-2.35}\cdot MdM\propto m_i^{-0.35} 
\end{equation}
for $m_i\ll m_{\rm max}$. Now, for the same example with $m_1=1$~M$_\odot$ and $m_2=100$~M$_\odot$, we have a ratio $M(>100\, {\rm M}_\odot)/M(>1\,{\rm M}_\odot)\sim100^{-0.35}\sim0.2$. The total masses in normal and heavy stars are thus comparable within a factor of $5$, and could become even more similar in case of a top-heavy IMF. In Eq.~\ref{crit}, the masses are multiplied with the inverse cube of their respective velocity dispersion, so that even a small difference in initial velocity dispersion could lead to a center that is strongly dominated by the massive component.

\subsection{Violent relaxation during star cluster mergers}\label{starclustermergers}

Mergers between star clusters can be crucial for the evolution and induce new violent relaxation processes. Here, we will focus mostly on the case of incomplete mixing. The latter is considered the most realistic scenario. However, we will also provide the case of complete mixing for comparison in Appendix~\ref{appendixcomplete}.

\subsubsection{Preliminary considerations}
We aim to extend our previous framework to the mergers of two star clusters A and B with individual stellar masses $m_1$ and $m_2$ with $m_2\gg m_1$. The number of normal stars in star cluster A is denoted as $N_{A1}$, the heavy stars in star cluster A as $N_{A2}$, the normal stars in star cluster B as $N_{B1}$, and the heavy stars in star cluster B as $N_{B2}$. We respectively define the masses of the populations as $M_{A1}=N_{A1}m_1$, $M_{A2}=N_{A2}m_2$, $M_{B1}=N_{B1}m_1$, $M_{B2}=N_{B2}m_2$. We assume that the components in each cluster are distributed following a Plummer sphere with Plummer radii $a_{A1}$, $a_{A2}$, $a_{B1}$, $a_{B2}$ and masses $M_{A1}$, $M_{A2}$, $M_{B1}$, $M_{B2}$. We assume the star clusters to be  virialized, as very likely each of them had time to go through a violent relaxation process before the time where they would merge. As a note of caution, we note that the Plummer spheres from the two stellar components in each cluster will be gravitationally interacting. We define a component virial parameter 
\begin{equation}
    q_{Jk}=2K_{Jk}/|W_{Jk}| \qquad J=A,B \qquad k=1,2
\end{equation} 
with $K_{Jk}$ the kinetic energy and 
\begin{equation} \label{WAB12}
    W_{Jk}=-3\pi GM_{Jk}^2/(32a_{Jk}).
\end{equation} 
Using this definition, we cannot assume that the q-parameters are equal to $1$, as in the virial equilibrium there will be a contribution to the kinetic term according to the gravitational interaction term. The condition of virial equilibrium for example for star cluster A would be\begin{equation}
2K_{A1}+2K_{A2}+W_{A1}+W_{A2}+W_{\rm int}=0,
\end{equation}
where the interaction term is given as\begin{eqnarray}
W_{\rm int}&=&-G\int\frac{M_{A1}(r)M_{A2}(r)}{r^2}dr\label{Wgrav}\\
&=&-GM_{A1}M_{A2}\int \frac{r^4dr}{(r^2+a_{A1}^2)^{3/2}(r^2+a_{A2}^2)^{3/2}}
\end{eqnarray}
with $M_{A1/2}$ the enclosed mass of component $1/2$. In case one of the clusters is much more compact than the other, for example $a_{A2}\ll a_{A1}$, we can approximate
\begin{eqnarray}
W_{\rm int}&=&-G\int\frac{M_{A1}(r)M_{A2}}{r^2}dr \\
&=&-GM_{A1}M_{A2}\int \frac{rdr}{(r^2+a_{A1}^2)^{3/2}}=-\frac{GM_{A1}M_{A2}}{a_{A1}}.\label{Wint}
\end{eqnarray}
Moreover, if we consider the scenario of $M_{A1}\gg M_{A2}$, the gravitational interaction term should  be negligible for component $1$ and we can  assume $q_{A1}\sim1$, while the cluster with lower mass will be dominated by the gravitational interaction term with the massive cluster. Comparing with $W_{A2}$, we find the interaction term to be relevant under the condition\begin{equation}
a_2\gtrsim \frac{3\pi}{32}\frac{M_{A2}}{M_{A1}}a_1
\end{equation}
This implies that only for very compact  distributions of the heavy component, the interaction term can be neglected. The ratio between both terms can then be approximated as\begin{equation}
\frac{W_{\rm int}}{W_{A2}}=\frac{32}{3\pi}\frac{M_{A1}}{M_{A2}}\frac{a_{A2}}{a_{A1}}.
\end{equation}

The above considerations show when the gravitational interaction
between the two components becomes important in the global energy
budget. We now also assess the virial equilibrium of the component $A2$, for which the general formulation of the scalar virial theorem (irrespective of the respective sizes of the clusters) is given as
\begin{equation}
2K_{A2}+W_{A2}+W_{A2,A1}=0,
\label{eq:virial_A2}
\end{equation}
with $W_{A2,A1}$  the gravitational binding energy of $A2$ due to the gravitational force exerted by $A1$,
\begin{equation}
W_{A2,A1}
=
-\int \rho_{A2}(\mathbf{r})\,
\mathbf{r}\cdot\nabla\Phi_{A1}(\mathbf{r})\,d^3r.
\label{eq:cross_virial}
\end{equation}
For spherical systems, this becomes
\begin{equation}
W_{A2,A1}
=
-4\pi G\int_0^\infty
\rho_{A2}(r) M_{A1}(r) r\,dr.
\label{eq:cross_virial_spherical}
\end{equation}
For the Plummer profiles considered here, we thus obtain
\begin{equation}
W_{A2,A1}
=
-\frac{3GM_{A1}M_{A2}}{a_{A2}^3}
\int_0^\infty
\frac{r^4\,dr}
{(r^2+a_{A1}^2)^{3/2}
 \left(1+r^2/a_{A2}^2\right)^{5/2}}.
\label{eq:cross_virial_plummer}
\end{equation}
Using the definition $q_{A2}=2K_{A2}/|W_{A2}|$, Eq.~(\ref{eq:virial_A2})
gives
\begin{equation}
q_{A2}
=
1+\frac{|W_{A2,A1}|}{|W_{A2}|}.
\label{eq:q_A2}
\end{equation}
Thus, $q_{A2}>1$ does not imply that component $A2$ is
supervirial with respect to the total gravitational potential. Rather,
it reflects the additional confinement provided by component $A1$,
as it is defined here relative to the self-gravity of
component $A2$. 

We can now also assess how the gravitational interaction energy depends on the relative size of the two Plummer spheres. A particularly simple limiting case is obtained when both components
have the same Plummer radius, $a_{A1}=a_{A2}=a_A$. The interaction
energy is then
\begin{equation}
W_{\rm int}
=
-\frac{3\pi}{16}
\frac{GM_{A1}M_{A2}}{a_A}.
\label{eq:Wint_equal}
\end{equation}
Since the two components have identical normalized spatial profiles,
the interaction contribution is shared symmetrically between their
virial terms, so that
\begin{equation}
W_{A2,A1}
=
\frac{1}{2}W_{\rm int}
=
-\frac{3\pi}{32}
\frac{GM_{A1}M_{A2}}{a_A}.
\label{eq:cross_equal}
\end{equation}
Comparing this with the self-gravitational energy of component $A2$,
we obtain
\begin{equation}
\frac{|W_{A2,A1}|}{|W_{A2}|}
=
\frac{M_{A1}}{M_{A2}},
\qquad
q_{A2}
=
1+\frac{M_{A1}}{M_{A2}}.
\label{eq:q_equal}
\end{equation}

Substituting $u=r/a_{A2}$ in Eq.~\ref{eq:cross_virial_plummer}, we obtain
\begin{equation}
W_{A2,A1}=
-\frac{3GM_{A1}M_{A2}a_{A2}^2}{a_{A1}^3}
\mathcal{I}\left(\frac{a_{A2}}{a_{A1}}\right)
\end{equation}
where we defined the function
\begin{equation}
\mathcal{I}(x)=
\int_0^\infty
\frac{u^4\,du}
{(1+u^2)^{5/2}(1+x^2u^2)^{3/2}},\label{Ix}
\end{equation}
with $x=a_{A1}/a_{A2}$. In the limit $a_{A2}\ll a_{A1}$, the integral scales approximately
as $\ln(a_{A1}/a_{A2})$, yielding
\begin{equation}
W_{A2,A1} \sim
-3\frac{GM_{A1}M_{A2}}{a_{A1}}
\left(\frac{a_{A2}}{a_{A1}}\right)^2
\ln\left(\frac{a_{A1}}{a_{A2}}\right).\label{A2A1}
\end{equation}
Comparing this expression with the self-gravitational energy of
component $A2$ using Eqs.~\ref{WAB12} and \ref{A2A1}, we find
\begin{equation}
\frac{|W_{A2,A1}|}{|W_{A2}|}
\sim
\frac{32}{\pi}
\frac{M_{A1}}{M_{A2}}
\left(\frac{a_{A2}}{a_{A1}}\right)^3
\ln\left(\frac{a_{A1}}{a_{A2}}\right).
\end{equation}
Thus, for a sufficiently compact massive component, the contribution
from component $A1$ to its virial balance becomes small and we recover
$q_{A2}\sim1$, provided that
\begin{equation}
\frac{M_{A1}}{M_{A2}}
\left(\frac{a_{A2}}{a_{A1}}\right)^3
\ln\left(\frac{a_{A1}}{a_{A2}}\right)
\ll 1.
\end{equation}

\subsubsection{Incomplete mixing} We assume in the case of incomplete mixing  that the energy exchange between the normal and the heavy stellar population is not efficient, so that their energies will be separately conserved. Into each population, energy is injected via the merger, and we assume that the injected energy will be proportional to the total mass in each population. In the following, we write down the equations for the low-mass population, but keeping in mind that the analogous equation will hold for the heavy stars as well. In a similar way, we take into account the gravitational energy due to the interaction of the two populations. We take $a_{A1}=a_{B1}/f_a$ and $a_{A2}=a_{B2}/f_a$, as well as $q_{A1}=q_{B1}=q_1$ and $q_{A2}=q_{B2}=q_2$, while $a_{f1}$ and $a_{f2}$ denote the final Plummer radii of the components 1 and 2 after the merger.

In the energy balance, we need to consider the total energy of the different mass components of the star clusters, which we can express in analalogy to Eq.~\ref{Etot} as\begin{equation}
E_{\rm tot A/B,1/2}=-\left(1-\frac{q_{A/B,1/2}}{2}\right)\frac{3\pi GM_{A/B,1/2}^2}{32a_{A/B,1/2}}.\label{EtotAB}
\end{equation}
In addition, we need to consider the gravitational interaction term for the components in the same clusters, which in the general form are given in Eq.~\ref{Wgrav}, as well as the orbital energy of the clusters, which we parametrize as
\begin{equation}
E_{\rm orbit}=-f_o\frac{GM_AM_B}{R_A+R_B},\label{Eorbit}
\end{equation}
with $f_0\sim1$ for a bound orbit, $f_0\sim-1$ for a hyperbolic orbit, and $f_0=0$ for a parabolic orbit. We recall $R_{A/B,1/2}=1.305a_{A/B,1/2}$ for a Plummer sphere. The energy balance can then be formally written as\begin{eqnarray}
&&K_{A1}+W_{A1}+W_{A1,A2}
+K_{B1}+W_{B1}+W_{B1,B2}
\nonumber\\
&&+E_{\mathrm{orbit}}
\frac{M_{A1}+M_{B1}}
{M_{A1}+M_{A2}+M_{B1}+M_{B2}}
\nonumber\\
&&=\frac{1}{2}\left(W_{1f}+W_{1,2,f}\right).
\end{eqnarray}
where $W_{1f}$ the corresponding gravitational potential energy of population 1 and $W_{1,
2f}$ the gravitational binding energy of population 1 due to populaton 2 (to be evaluated via Eq.~\ref{eq:cross_virial}). We further assume virial equilibrium for the final configuration. The sum of kinetic plus gravitational energy for one component is evaluated via Eq.~\ref{EtotAB}, where for the final configurations we assume that virialization has been achieved. The terms corresponding to the gravitational interaction energy are described via Eq.~\ref{eq:cross_equal}, and the term corresponding to the injected orbital energy via Eq.~\ref{Eorbit}. For the last term, we further assume that the fraction of that energy that enters the relevant energy balance for population 1 corresponds to its respective mass fraction. Our overall energy balance is then
\begin{eqnarray}
&&-\frac{3\pi G}{32}\left(1-\frac{q_{A1}}{2}\right)
\frac{M_{A1}^2}{a_{A1}}
-\frac{3\pi G}{32}\left(1-\frac{q_{B1}}{2}\right)
\frac{M_{B1}^2}{a_{B1}}
\\
&&-\frac{3GM_{A1}M_{A2}a_{A1}^2}{a_{A2}^3}
\mathcal{I}\left(\frac{a_{A1}}{a_{A2}}\right)
-\frac{3GM_{B1}M_{B2}a_{B1}^2}{a_{B2}^3}
\mathcal{I}\left(\frac{a_{B1}}{a_{B2}}\right)
\nonumber\\
&&-\frac{f_0G(M_{A1}+M_{A2})(M_{B1}+M_{B2})}
{1.305(a_{A1}+a_{B1})}
\frac{M_{A1}+M_{B1}}
{M_{A1}+M_{A2}+M_{B1}+M_{B2}}
\nonumber\\
&&=-\frac{3\pi G(M_{A1}+M_{B1})^2}{64a_{f1}} \nonumber \\
&&-\frac{3G(M_{A1}+M_{B1})(M_{A2}+M_{B2})a_{f1}^2}
{2a_{f2}^3}
\mathcal{I}\left(\frac{a_{f1}}{a_{f2}}\right).\label{fullbalance}\nonumber
\end{eqnarray}
We keep in mind that the same equation holds for the second population, just interchanging indices 1 and 2. We consider again the limit of clusters with equal masses, namely $M_{A1}=M_{B1}=M_1$ and $M_{A2}=M_{B2}=M_2$. For simplicity, we first assume that the two stellar populations within each progenitor cluster have identical initial Plummer radii, $a_{A1}=a_{A2}$ and $a_{B1}=a_{B2}$. The radii of the two populations in the merger remnant, however, are allowed to differ. The equation then becomes
\begin{eqnarray}
&&\frac{1}{a_{f1}}\left[
1+\frac{32}{\pi}\frac{M_2}{M_1}
\left(\frac{a_{f1}}{a_{f2}}\right)^3
\mathcal{I}\left(\frac{a_{f1}}{a_{f2}}\right)
\right]=\frac{1+f_a^{-1}}{2a_{A1}}
\left(1-\frac{q_1}{2}+\frac{M_2}{M_1}\right)
\nonumber\\
&&+\frac{16f_0}{3.915\pi a_{A1}}
\frac{1+M_2/M_1}{1+f_a}.\label{incompleteenergy}
\end{eqnarray}
We further assume that both stellar populations contribute equally to the total mass, $M_1=M_2$. If we further assume that the two populations have identical final Plummer radii, $a_{f1}=a_{f2}$, we can use $\mathcal{I}(1)=\pi/32$ to obtain an explicit expression for the final radius,
\begin{equation}
a_{f1}=
\frac{2a_{A1}}
{\frac{1+f_a^{-1}}{2}(2-\frac{q_1}{2})
+\frac{32f_0}{3.915\pi(1+f_a)}}.
\end{equation}
For initially virialized progenitor clusters, we have
$q_1=2$ in this case. In particular, for a parabolic merger
($f_0=0$), we obtain
$a_{f1}=4f_a a_{A1}/(1+f_a)$.
For $f_a=1$, this reduces to $a_{f1}=2a_{A1}$,
as expected for the merger of two identical virialized clusters.

If instead we consider Eq.~\ref{incompleteenergy} in the limit $M_2\ll M_1$, we obtain the expression
\begin{equation}
a_{f1}=
\frac{2a_{A1}}
{\frac{1+f_a^{-1}}{2}(2-q_1)
+\frac{32f_0}{3.915\pi(1+f_a)}}.
\label{afLimmit}
\end{equation}
In case of negligible orbital energy, $f_0=0$, we find
\begin{equation}
a_{f1} =
\frac{2f_a}{1+f_a}
\left(1-\frac{q_1}{2}\right)^{-1}a_{A1}.\label{afsmall}
\end{equation}

\begin{figure}
\includegraphics{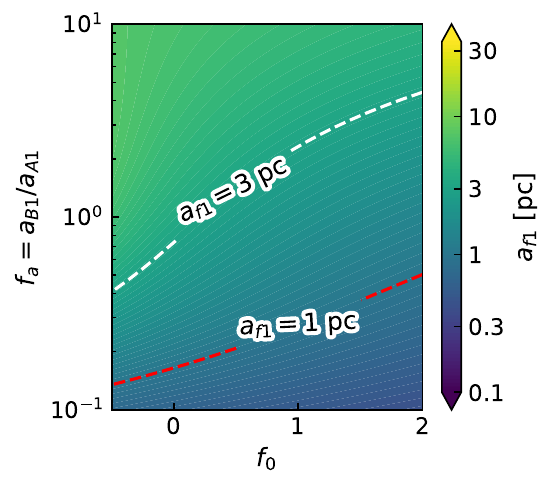}
\caption{The final Plummer radius of component $1$ as a function of the parameter $f_a$ characterizing the ratio between the Plummer spheres of the two different clusters and the parameter $f_0$ corresponding to the normalization of the orbital energy. The Plummer radius is calculated via Eq.~\ref{afLimmit} in the limit $M_2\ll M_1$.  We assume here that initially $a_{A1}=3$~pc and $q_1=0.3$.}
\label{afM}
\end{figure}

We note that particularly when $f_a\ll1$, it will play an important role in determining the new Plummer radius and we obtain $a_{f1}\sim f_aa_{A1}$. The merger with a compact cluster thus tends to lead to a further compactification. We provide the resulting Plummer radius $a_{f1}$ as a function of $f_a$ and $f_0$ in Fig.~\ref{afM}. It shows in particular that $a_{f1}$ decreases both with $f_0$ while it increases with $f_a$, though the dependence is non-linear. For initial values of $a_{A1}=3$~pc and $q_1=0.3$, more compact radii can be obtained preferentially for values of $f_a\ll1$, so when one of the merging clusters is much more compact than the other one.

The corresponding energy balance for the massive stellar
population can be obtained by interchanging indices 1 and 2
in Eq.~\ref{fullbalance}. Adopting the same assumptions concerning the
masses and initial Plummer radii of the progenitor clusters,
we obtain

\begin{eqnarray}
&&-\frac{3\pi G M_2^2}{32a_{A2}}
\left(1+\frac{1}{f_a}\right)
\left(1-\frac{q_2}{2}+\frac{M_1}{M_2}\right)-\frac{f_0G(M_1+M_2)M_2}
{1.305a_{A2}(1+f_a)}
\nonumber\\
&&=-\frac{3\pi GM_2^2}{16a_{f2}}
-6GM_1M_2\frac{a_{f2}^2}{a_{f1}^3}
\mathcal{I}\left(\frac{a_{f2}}{a_{f1}}\right).\label{massivecomponent}
\end{eqnarray}
We now consider the limiting case $M_2\ll M_1$, where
the massive stellar population contributes only a small
fraction of the total cluster mass. Since the progenitor
clusters are initially virialized and the two populations
have identical initial Plummer radii, Eq.~\ref{eq:q_equal} implies
$q_2=1+M_1/M_2$. Consequently, the combination
appearing in Eq.~\ref{massivecomponent} becomes
\begin{equation}
M_2^2\left(1-\frac{q_2}{2}+\frac{M_1}{M_2}\right)
=\frac{M_2(M_1+M_2)}{2}.
\end{equation}
Thus, the initial energy of population 2 remains
proportional to $M_1M_2$ in this limit, reflecting
its gravitational interaction with the dominant
population 1. The final cross-virial term is of the
same order and must therefore also be retained.
In contrast, the final self-gravitational energy
of population 2 is proportional to $M_2^2$ and
can be neglected to leading order, provided its
final Plummer radius remains finite.  This gives us the following expression:
\begin{equation}
6\frac{a_{f2}^2}{a_{f1}^3}
\mathcal{I}\left(\frac{a_{f2}}{a_{f1}}\right)
=\frac{3\pi}{64a_{A2}}
\left(1+\frac{1}{f_a}\right)
+\frac{f_0}{1.305a_{A2}(1+f_a)}.\label{afunequalmass}
\end{equation}
Another interesting case is the limit of negligible orbital
energy, $f_0=0$. For initially virialized progenitors
and $M_2\ll M_1$, we have $q_1\simeq 1$, so Eq.~\ref{afsmall}
gives $a_{f1}=4f_a a_{A1}/(1+f_a)$.
Substituting this expression into Eq.~\ref{afunequalmass}, and using
$a_{A1}=a_{A2}$, we obtain
\begin{equation}
\left(\frac{a_{f2}}{a_{f1}}\right)^2
\mathcal{I}\left(\frac{a_{f2}}{a_{f1}}\right)
=\frac{\pi}{32}.
\end{equation}
The equation admits the solution $a_{f2}=a_{f1}$,
since $\mathcal{I}(1)=\pi/32$. Numerical evaluation
also yields a second solution,
$a_{f2}/a_{f1} \simeq 1.435$. For the equal-radius branch, we find
\begin{equation}
a_{f2}=\frac{4f_a}{1+f_a}a_{A2}.\label{mergersimplified}
\end{equation}
The result shows that mergers with more compact
progenitor clusters lead to more compact remnants.
However, under the present assumptions, the merger
does not generate additional mass segregation between
the two stellar populations.

\section{Build up of a massive compact star cluster
through violent relaxation}\label{massivecluster}
We now consider a simple toy model for the assembly of a
massive stellar system through the merger of $N_c$ initially
virialized star clusters. For simplicity, we assume that
all progenitor clusters have identical masses $M_c$ and
Plummer radii $a_c$, and that the merger remnant can again
be described by a virialized Plummer sphere with total
mass $N_c M_c$ and Plummer radius $a_f$. We further
neglect mass loss during the merger process.

The initial internal energy of each progenitor cluster
is given by the virial theorem, corresponding to
\begin{equation}
E_c=-\frac{3\pi G M_c^2}{64a_c}.
\label{eq:multi_cluster_energy}
\end{equation}
In addition to the internal energies, we need to
account for the orbital energy of the clusters prior
to their merger. In analogy to Eq.~(\ref{Eorbit}),
we parametrize the orbital energy through a
dimensionless parameter $f_0$, with $f_0=0$
corresponding to vanishing orbital energy and
$f_0>0$ to an initially bound configuration.

We introduce a characteristic interaction scale
\begin{equation}
R_{\rm int}=\xi(N_c)a_c,
\label{eq:multi_interaction_radius}
\end{equation}
with $\xi(N_c)$ being some increasing function of $N_c$. This expression allows the characteristic scale to increase with the number of cluster and thus the total amount of mass involved in the interaction. To estimate this dependence, we compare the self-gravitational acceleration within one progenitor cluster with the tidal
acceleration produced by the assembling system. The
characteristic self-gravitational acceleration of a Plummer
cluster at $r=a_c$ is
\begin{equation}
g_{\rm self}=\frac{GM_c}{2^{3/2}a_c^2},
\label{eq:multi_self_gravity}
\end{equation}
while the differential tidal acceleration due to an effective
perturbing mass $M_{\rm pert}$ at a separation $d$ is
\begin{equation}
g_{\rm tidal}\sim\frac{2GM_{\rm pert}a_c}{d^3}.
\label{eq:multi_tidal_acceleration}
\end{equation}
Here we have adopted the distant-tide approximation and
treated the perturbing mass as centrally concentrated.
Equating Eqs.~(\ref{eq:multi_self_gravity}) and
(\ref{eq:multi_tidal_acceleration}) yields
\begin{equation}
d_{\rm tidal}\sim 2^{5/6}
\left(\frac{M_{\rm pert}}{M_c}\right)^{1/3}a_c.
\label{eq:multi_tidal_radius}
\end{equation}
For an individual cluster interacting with the other
$N_c-1$ progenitors, we adopt the effective perturbing
mass $M_{\rm pert}=(N_c-1)M_c$, which approaches the
total assembled mass $N_cM_c$ when $N_c\gg1$. We thus
obtain the mass-dependent interaction scale
\begin{equation}
R_{\rm int}=\xi_0(N_c-1)^{1/3}a_c,
\qquad \xi_0=2^{5/6}\simeq1.78,
\label{eq:multi_interaction_radius_N}
\end{equation}
for $N_c\geq2$. We thus obtain
$\xi(N_c)=\xi_0(N_c-1)^{1/3}$. As the point-mass tidal approximation becomes
inaccurate when the perturbing mass is spatially extended
or the separation is comparable to the cluster radius, the estimate in
Eq.~(\ref{eq:multi_interaction_radius_N}) should therefore
be regarded as indicative rather than as a precise
threshold. 

For a system of $N_c$ clusters, the number of
distinct cluster pairs is
\begin{equation}
N_{\rm pair}=\frac{N_c(N_c-1)}{2}.
\label{eq:multi_number_pairs}
\end{equation}
Assuming that each pair contributes approximately
the same net orbital energy, characterized by the
interaction scale in
Eq.~(\ref{eq:multi_interaction_radius_N}), we obtain
\begin{equation}
E_{\rm orbit}=
-f_0\frac{GN_c(N_c-1)M_c^2}
{2\xi_0(N_c-1)^{1/3}a_c}
=-f_0\frac{GN_c(N_c-1)^{2/3}M_c^2}
{2\xi_0a_c},
\label{eq:multi_orbital_energy}
\end{equation}
for $N_c\geq2$, with $E_{\rm orbit}=0$ for $N_c=1$.
The factor $N_c(N_c-1)/2$ avoids double-counting cluster
pairs. Here $f_0=0$ corresponds to vanishing net orbital
energy, whereas $f_0>0$ corresponds to a bound initial
configuration. This prescription remains an effective
approximation, since the actual orbital energy depends on
the distribution of pair separations and relative
velocities. In particular, the onset of tidal interactions
does not by itself determine the net orbital energy. Assuming conservation of the total energy,
the sum of the initial internal energies and
the orbital energy must equal the final
binding energy of the merger remnant:
\begin{eqnarray}
&&N_c E_c+E_{\rm orbit}
=-\frac{3\pi G(N_cM_c)^2}{64a_f}.
\label{eq:multi_energy_conservation}
\end{eqnarray}
Inserting Eqs.~(\ref{eq:multi_cluster_energy})
and (\ref{eq:multi_orbital_energy}), we find
\begin{equation}
-\frac{3\pi G N_cM_c^2}{64a_c}
-\frac{f_0GN_c(N_c-1)^{2/3}M_c^2}{2\xi_0a_c}
=-\frac{3\pi GN_c^2M_c^2}{64a_f}.
\label{eq:multi_energy_balance}
\end{equation}
Solving for the final Plummer radius yields
\begin{equation}
a_f=
\frac{N_ca_c}
{1+\frac{32f_0(N_c-1)^{2/3}}{3\pi\xi_0}}.
\label{eq:multi_final_radius}
\end{equation}
Equation~(\ref{eq:multi_final_radius}) illustrates
how the final radius depends on the number of merging
clusters and their initial orbital energy. For $f_0=0$,
we recover $a_f=N_ca_c$, corresponding to a remnant whose
radius increases linearly with the number of progenitors.
For $f_0>0$, the additional negative orbital energy leads
to a more compact final configuration. Unlike a model with
a fixed interaction radius, the present prescription does
not predict a constant final radius as $N_c$ increases.
For any fixed $f_0>0$ and $N_c\gg1$,
Eq.~(\ref{eq:multi_final_radius}) approaches
\begin{equation}
a_f\simeq\frac{3\pi\xi_0}{32f_0}N_c^{1/3}a_c,
\label{eq:multi_radius_limit_N}
\end{equation}
so that $a_f\propto M_{\rm tot}^{1/3}$ for
$M_{\rm tot}=N_cM_c$. Consequently, the characteristic
mean density $M_{\rm tot}/a_f^3$ approaches a constant
for fixed $M_c$, $a_c$, and $f_0>0$. The denominator in
Eq.~(\ref{eq:multi_final_radius}) remains positive for
all $N_c\geq1$ when $f_0\geq0$, including $f_0=1$.
This relation depends on the assumption that
the effective perturbing mass scales with the total
assembled mass and that a single characteristic
interaction scale adequately describes the cluster pairs.

For negative values of $f_0$, corresponding to positive
initial orbital energy in our parametrization, the
denominator in the expression for the final radius
decreases with increasing $N_c$. It vanishes at
\begin{equation}
N_{\rm crit}
=
1+
\left(
\frac{3\pi\xi_0}{32|f_0|}
\right)^{3/2},
\end{equation}
where the total energy of the system becomes zero
and the predicted equilibrium radius diverges.
For $N_c>N_{\rm crit}$, the total energy becomes
positive, and the assumed bound final Plummer
configuration is no longer energetically possible.

This behavior should not necessarily be interpreted
as the physical disruption of an already existing
central cluster, but rather as a limitation of the
assumption that all $N_c$ clusters eventually merge
into a single bound object. In particular, if $f_0$
is negative for the cluster system as a whole, its
positive orbital energy implies that the initial
configuration is globally unbound when the orbital
energy is considered separately from the internal
binding energies of the individual clusters.
Under these circumstances, many clusters may escape
without merging, especially if their relative
velocities are sufficiently large to prevent
efficient gravitational capture. Individual mergers
may nevertheless occur between clusters with
favorable relative velocities and trajectories,
potentially forming a bound central subsystem.
Thus, for increasingly negative orbital energies
or sufficiently large $N_c$, the number of clusters
actually participating in mergers may be
substantially smaller than the initial $N_c$.
The divergence of the analytical expression
therefore indicates the breakdown of the
complete-merger assumption rather than predicting
the destruction of the central cluster.

\begin{figure}[!h]
    \centering
\includegraphics{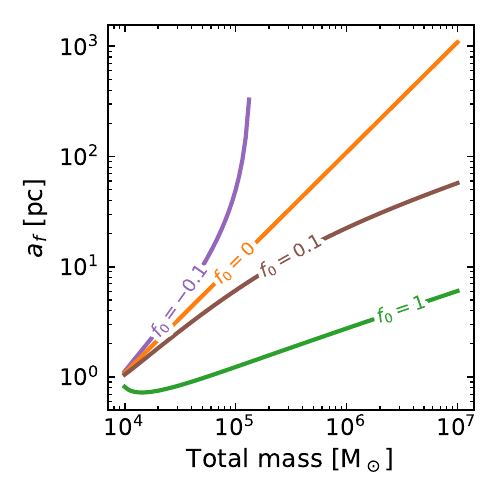}
    \caption{ Resulting radius of the massive cluster formed through violent relaxation (Eq. \ref{eq:multi_final_radius}) as function of the total mass of the system assuming $m_{\rm cav}=9.2\times10^3$\,M$_\odot$. In here, we adopted $a_c=1$\,pc and different values for $f_0$. Specifically, $f_0=[-0.1,0,0.1,1]$ in purple, orange, brown, and green, respectively. We note that a globally negative value of $f_0$ is a somewhat artificial extreme case, as on the one hand it means that the clusters should be unbound, while nonetheless the model assumes that many of them would merge. The injection of too much positive energy into the cluster then leads to a significant expansion. In this regime, the model is however not fully adequate as it is likely that also mass loss and partial disruptions of the cluster should be taken into account.}
    \label{fig:finalSize}
\end{figure}
The resulting evolution is shown in Fig.~\ref{fig:finalSize}. For the toy model considered here, the final radius increases as a function of the total mass. In case of the mergers of gravitationally bound clusters (positive $f_0$), we obtain more compact configurations, while the cluster becomes wider for larger values of $f_0$. In case of positive $f_0$, then after too many mergers the configuration is no longer stable due to the substantial energy input. We note that this is also a rather extreme case, as in case of a global positive $f_0$, the configuration is overall unbound, and it is unlikely that too many mergers would subsequently happen. If it was the case, it is very likely that not all of the clusters were bound, but part of the stars could be ejected from the cluster, affecting the energy balance. It is thus very likely that the model would require further improvements in this regime. 

Overall, one of the most favorable pathways to form a very dense cluster and subsequently a massive black hole is  the violent relaxation of a cold cluster, preferably in the presence of a massive stellar population, as outlined in section~\ref{coldcollapsemassive}. In the presence of a sufficiently high number of massive stars, the heavy central component may decouple from the general cluster during violent relaxation. At this stage, it is likely to exhibit densities where runaway collisions can occur, as suggested by  \citet{Escala2021, Vergara2023, Vergara2025a, Vergara2025b}.

We consider a massive star cluster with $M_c=10^7$~M$_\odot$ as an example. As discussed in section~\ref{coldcollapsemassive}, under the assumption of a Salpeter IMF, we then have a total mass in normal stars of $M_1=8\times10^6$~M$_\odot$ and a total mass in massive stars of $M_2=2\times10^6$~M$_\odot$. We assume the normal component to have a Plummer radius $a_1=3$~pc, while we take the massive component to be more concentrated with $a_2=1$~pc. If both Plummer distributions were virialized, they had velocity dispersions of $109$~km/s and
$94$~km/s, respectively. We assume that the normal stellar component is somewhat subvirial, with $q_{1i}\sim0.3$, while for the massive component, we assume it to form from the high-density gas material that is relatively at rest with respect to the center of mass, adopting a virial parameter $q_{2i}=0.15$. These virial ratios imply initial velocity dispersions of $\sigma_{1i}=59.7$~km/s and $\sigma_{2i}=36.4$~km/s. Following Eq.~\ref{beta}, this gives final velocity dispersions after violent relaxation of $\sigma_{1f}\sim70.6$~km/s and $\sigma_{2f}\sim43.1$~km/s.

We use Eq.~\ref{crit} as the condition for whether the more massive component will dynamically separate from the less massive one. We find indeed that\begin{equation}
M_s\sigma_{2f}^{-3}\sim24.9\frac{M_\odot}{{\mathrm{km/s}}}>M_1\sigma_{1f}^{-3}\sim22.7\frac{M_\odot}{{\mathrm{km/s}}},
\end{equation}
implying the density of the massive stars to be larger in the center and suggesting a dynamical decoupling from the other component. While we note that here the condition is only marginally fulfilled, the effect is likely to become considerably more pronounced in case of a top-heavy IMF. Assuming the massive stars to be concentrated in a region of less than a parsec, we have a number density of $\sim20\,000$~pc$^{-3}$. The relaxation time in this central region, according to Eq.~\ref{relax}, corresponds to $4.5\times10^6$~years. We therefore expect this region to go through efficient core collapse as described for example by \citet{Vergara2025a}.

\section{Discussion and conclusions}\label{conclusions}

\begin{table*}
\caption{Summary of the main limiting cases considered in this work.
Here, $M_1$ and $M_2$ denote the masses of the normal and massive
stellar components, respectively, $a_1$ and $a_2$ their Plummer
radii, $q_i$ their virial parameters, and $f_0$ parametrizes the
orbital energy in cluster mergers. For multiple mergers, $N_c$ is
the number of progenitors, $a_c$ their initial Plummer radius,
and $\xi_0$ the interaction-scale parameter. The dimensionless parameter $f_a$ is given as $f_a=a_{B1}/a_{A1}=a_{B2}/a_{A2}$ for the star clusters A and B, and the indices 1 and 2 refer to the light and heavy stellar component. It is implicitly assumed here that the ratio is the same for the light and heavy component. }
\label{tab:limiting_cases}
\centering
\begin{tabular}{p{3.2cm} p{4.6cm} p{5.2cm} p{3cm}}
\hline\hline
Process & Limiting case & Main result & Favors black hole formation? \\
\hline

Cold collapse,
single component
&
Subvirial initial cluster,
$q<1$
&
$\displaystyle
a_f=\frac{a_i}{2-q}
$ & yes
\\[2ex]

Cold collapse,
two components
&
Complete mixing &

$a_f=\frac{1}{2}\frac{(M_1+M_2)^2a_1a_2}{(1-\frac{q_1}{2})M_1^2a_2+(1-\frac{q_2}{2})M_2^2a_1},$

& possibly
\\[2ex]

Cold collapse,
two components
&
Incomplete mixing;
inefficient energy exchange
&
$\displaystyle
M_2\sigma_{2f}^{-3}
>
M_1\sigma_{1f}^{-3}
$ & yes
\\[2ex]

Merger of two clusters
&
Incomplete mixing,
$M_2\ll M_1$, $f_0\simeq0$;
light component
&
$\displaystyle
a_{f1}
=
2\left(1-\frac{q_1}{2}\right)^{-1}
(1+f_a^{-1})^{-1}a_{A1}
$ & no
\\[2ex]

Merger of two clusters
&
Incomplete mixing,
$M_2\ll M_1$, $f_0>0$;
heavy component
&
Eq.~\ref{afunequalmass}; see Eq.~\ref{mergersimplified} for a simplified version.

& no 
\\[2ex]

Multiple cluster mergers
&
$N_c$ identical virialized
clusters; $\xi(N_c)=\xi_0(N_c-1)^{1/3}$
&
$\displaystyle
a_f=\frac{N_ca_c}{1+\frac{32f_0}{3\pi\xi_0}(N_c-1)^{2/3}}
$ & no
\\[2ex]

Cluster merger
&
Complete mixing,
$a_{A2}\ll a_{A1}$,
$M_1=M_2$, $f_0=0$, $f_a=1$
&
$\displaystyle
a_f\simeq 8\left[\frac{1}{a_{A1}}+\frac{1}{a_{A2}}+
\frac{32}{3\pi a_{A1}}\right]^{-1}
$
& no
\\

\hline
\end{tabular}\label{summarytable}
\end{table*}

Large stellar systems, including galaxies or galaxies in formation,  likely consist of sub-structures, reflecting the initial hierarchical formation process. Star clusters forming from cold initial conditions will go through an initial phase of violent relaxation, leading to a compactification of the cluster and bringing them into approximate virial equilibrium. The different scenarios and limiting cases we analyzed here are summarized in Table~\ref{summarytable}.

In this manuscript, we have shown that hierarchical stellar systems in the early Universe may evolve on short timescales via purely stellar dynamical processes. While the 2-body relaxation time may formally be large, stellar systems likely have a substructure consisting of star clusters, which each individually will move approximately as a point particle. For such a system of star clusters, we can define a cluster relaxation time as introduced in section~\ref{collisional}, which is considerably reduced compared to the 2-body relaxation time by a factor corresponding to the typical number of stars per cluster.

Violent relaxation can be particularly interesting in star clusters that include a relevant component of high-mass stars. As the process of violent relaxation is likely to remain incomplete given the short timescale over which the gravitational potential keeps changing, it is quite likely that the light and heavy components relax separately without efficiently exchanging energy during the process. In such a case, the massive stellar component may become more centrally concentrated and can potentially decouple from the main star cluster. We derived a quantitative criterion for this process in section~\ref{coldcollapsemassive}. This type of decoupling has some similarity but also important differences compared to the Spitzer instability \citep{Spitzer1987}. While the final result in both cases is a decoupling of the heavy and light stellar component, for the Spitzer instability this process is driven by gravitational encounters, while here we consider a completely collisionless process due to violent relaxation. It is important to notice that in the collisionless limit, such a mass segregation can only occur if there is already a difference in the initial phase space structure of the light and the heavy component.

We have further investigated violent relaxation during the mergers of star clusters. We find in particular that mergers with particularly dense clusters can help to form a more compact system, though at least typically the merger induces a certain expansion both for the light and the massive stellar component.

In section~\ref{massivecluster}, we considered the build-up of massive star clusters via multiple collisions, leading to a significant increase of the radius with the number of collisions. The formation of massive but compact clusters should thus preferably occur via no mergers or a moderate number of mergers from gravitationally bound clusters, while multiple mergers will lead to massive but more extended stellar systems. In the first case, where a massive compact systems forms after few or no mergers, core collapse may induce runaway collisions, leading to the formation of a very massive central object \citep[e.g.,][]{Portegies2002,Reinoso2018,  Escala2021, Vergara2023, Vergara2024,Rantala2025, Rantalla2025b}. 
\\ \indent
In the alternative case, when a massive stellar system builds up via multiple mergers, it is less likely that the final system itself will be able to go through core collapse on a short enough timescale. This does not necessarily exclude the possibility to have intermediate-mass black holes in such a system, but rather depends on the previous evolution of the star clusters from which it forms. In particular, their progenitor clusters may have gone through a cold collapse inducing violent relaxation, and potentially in some of them the massive stars subsequently could have formed a dynamically decoupled core, as discussed in section~\ref{coldcollapsemassive}, which subsequently may produce a central massive object via core collapse. Nonetheless, at least the formation of a central massive black hole is likely more feasible via core collapse in a massive compact system, as opposed to the core collapse in a lower-mass progenitor cluster. 
\\ \indent
In summary, we emphasize that sub-structure in massive stellar systems as well as violent relaxation events due to cold collapse and mergers provide rapid pathways for the evolution of such stellar systems. This can be relevant both for the formation of massive bulges and ellipticals, as proposed by \citet{Tremaine1986, akiyama1989, Bournaud2011} and \citet{Dekel2022}. The cold collapse leads to more compact configurations, while mergers potentially to more extended ones, unless the clusters are strongly bound. The formation of very massive objects can be facilitated via cold collapse of very massive clusters, particularly in the presence of a significant high mass stellar component. The latter may dynamically decouple as a result of violent relaxation, and subsequently go through core collapse.

Throughout this manuscript, we have neglected the possible effects of gas. The latter could further aid in the dissipation of energy, thereby leading to a further compactification of the systems considered here, and could also aid in the formation of massive objects via dynamical friction and through gas accretion. A treatment of these effects is beyond the scope of this work; nonetheless we can conclude that the potential for rapid evolution exists even within the purely stellar dynamical limit.

Also, in the context of cluster mergers, we have here focused on the limiting case of clusters that do not yet include a central massive black hole. It is conceivable that such a massive black hole could form at an earlier stage, for example as a result of core collapse in a dense massive cluster, particularly when the core collapse timescale is shorter than the timescale for hierarchical cluster mergers. The evolution of realistic star clusters will thus be determined both by their individual evolution when/if they are unperturbed as well as the properties of their environment. While mergers of clusters increase the available mass of the system, our analysis here shows that also the corresponding energy injection is highly relevant. Once a massive black hole is present within a star cluster, we speculate that it could have very similar effects during violent relaxation as a centrally concentrated population of quite massive stars (a limiting case we investigated here), thus potentially producing more compact systems during violent relaxation, which may subsequently favor the feeding of the black hole via tidal disruption events.

\begin{acknowledgements}
DRGS, ML, TZ and LP acknowledge support from the MUR projects FIS-2024-01621 DAWN. DRGS  thanks for support from the Alexander von Humboldt - Foundation, Bonn, Germany. Furthermore, AE, DRGS and MCV gratefully acknowledges funding via the ANID BASAL project FB21003.      
This material is based upon work supported by Tamkeen under the NYUAD Research Institute grant CG014.
ML gratefully acknowledges support from ANID/DOCTORADO BECAS CHILE 72240058.
MAL
thanks the UAEU for funding via UPAR grant No.
G00005454.

\end{acknowledgements}


%

\appendix
\nolinenumbers

\section{The role of the Coloumb parameter with the different massive population of stars in a cluster} \label{app:columb}

The analysis of the different massive populations in a newly formed star cluster is fundamental in our work. Thus, we also briefly describe the different values of the Coulomb parameter $\Lambda$ in the case of different stellar mass spectra or multiple number of populations. The Coulumb parameter is an empirical value, and it has been extrapolated from simulations \citep[both from Montecarlo and N-body direct simulations, see e.g.,][]{giersz1994a,freitagandbenz2001}. Most of its values span between 0.01 to 0.40, which depend on (i) the mass spectra of the population(s) and (ii) the presence of multiple mass population (e.g., a population of planets and one of stars). The main values used in the literature are the following:
\begin{enumerate}
    \item The value of 0.4 \citep{Spitzer1987,giersz1994a} is considered the "standard" value, as it is constructed from a typical \cite{Kroupa2001} initial mass function, for star cluster model between $10^4$ and $10^5$ stars. This is typical for a single stellar population star cluster and quite popular in large $N$-body direct simulations or Monte Carlo methods.
\item The value of  0.01-0.02 is instead used for multiple mass population. This value is often considered valid if the Spitzer instability criteria \cite{Spitzer1969} is not confirmed, thus $M_2 \ll M_{\rm 1,c}$, where $M_2$ is the less massive population and $M_{\rm 1,c}$ is the more massive population. In this scenario, the less massive population is thermically irrelevant, and the less massive particles do not participate in the process of mass segregation \citep{flamminidotti2025, flamminidotti2026}.
\item The value of 0.11 \citep{giersz1994a,joshirasiopz2000, freitagandbenz2001}is used in the absence of a stellar mass spectra in the cluster (i.e., all the masses have the same value in the cluster). This value is typically encountered in old star clusters (with ages $>$ Gyr) or in toy models.
\end{enumerate}
There are additional values in the literature, but these are the most recognized and cited by the community.

\section{Cold collapse of initially sub-virial King and Dehnen models}\label{King}
In the following, we generalize the results from section~\ref{coldsimple} to the models of \citet{King1962} and \citet{Dehnen1993}.

\subsection{King models}
We  consider the evolution of an initially sub-virial system whose
density distribution is described by a King model \citep{King1962, Binney2009}. King models constitute a
family of truncated equilibrium distribution functions that can be
characterized by the dimensionless central potential $W_0$, or equivalently
by a concentration parameter. In addition to setting the physical scale
through the King radius $r_{\rm K}$, the value of $W_0$ determines the
dimensionless structure of the system. For a King model of total mass $M$, we write the gravitational potential
energy as
\begin{equation}
    W = -\kappa(W_0)\frac{G M^2}{r_{\rm K}},
\end{equation}
where $\kappa(W_0)$ is a dimensionless structural coefficient that depends
only on the concentration of the King model. We characterize the initial
dynamical state through the virial ratio
\begin{equation}
    q_i = \frac{2K_i}{|W_i|},
\end{equation}
such that $q_i=1$ corresponds to virial equilibrium, while $q_i<1$
describes a sub-virial configuration. An initially sub-virial King model
can be constructed by preserving its initial density
distribution while reducing the velocities relative to the
corresponding equilibrium King model.

The initial total energy is then
\begin{equation}
    E_i = K_i + W_i
        = -\left(1-\frac{q_i}{2}\right)|W_i|=
    -\left(1-\frac{q_i}{2}\right)
    \kappa(W_{0,i})\frac{G M^2}{r_{{\rm K},i}}.
\end{equation}

We assume that the subsequent cold collapse and violent relaxation lead to a new configuration that can be described by a virialized King model. For the final system,
\begin{equation}
    E_f = \frac{W_f}{2}
        = -\frac{1}{2}\kappa(W_{0,f})
          \frac{G M^2}{r_{{\rm K},f}}.
\end{equation}
Neglecting mass loss and assuming that the total energy is conserved during
violent relaxation, $E_i=E_f$, we obtain
\begin{equation}
    r_{{\rm K},f}
    =
    \frac{\kappa(W_{0,f})}{\kappa(W_{0,i})}
    \frac{r_{{\rm K},i}}{2-q_i}.
\label{eq:king_final_radius}
\end{equation}
Equation~(\ref{eq:king_final_radius}) separates two effects. The factor
$(2-q_i)^{-1}$ describes the contraction required by energy conservation
when an initially sub-virial system reaches virial equilibrium, while the
ratio $\kappa(W_{0,f})/\kappa(W_{0,i})$ accounts for changes in the
dimensionless structure of the system during violent relaxation.

In the limiting case where the King concentration remains unchanged,
$W_{0,f}=W_{0,i}$, the structural coefficients cancel and
Eq.~(\ref{eq:king_final_radius}) reduces to
\begin{equation}
    r_{{\rm K},f}
    =
    \frac{r_{{\rm K},i}}{2-q_i},
\end{equation}
which is the result we found for the Plummer model in Eq.~\ref{afcold}.  

\subsection{Dehnen models}
We  consider the Dehnen density profile \citep{Dehnen1993, Binney1987}, given by
\begin{equation}
\rho(r)=\frac{(3-\gamma)M}{4\pi}
\frac{a}{r^\gamma(r+a)^{4-\gamma}},
\end{equation}
where $M$ is the total mass, $a$ is the scale radius,
and $0\leq\gamma<3$ describes the inner density slope.
The enclosed mass is
\begin{equation}
M(r)=M\left(\frac{r}{r+a}\right)^{3-\gamma},
\end{equation}
and the gravitational potential energy is
\begin{equation}
W=-4\pi G\int_0^\infty \rho(r)M(r)r\,dr
=-\frac{GM^2}{2a(5-2\gamma)},
\end{equation}
where the latter expression is valid for $\gamma<5/2$. We assume an initially sub-virial configuration with
virial ratio $q_i=2K_i/|W_i|$, which undergoes
violent relaxation and subsequently reaches virial
equilibrium. The initial total energy is
\begin{equation}
E_i=-\frac{(2-q_i)GM^2}
{4a_i(5-2\gamma_i)},
\end{equation}
taking $\gamma_i$ as the initial value of the inner slope. For the final configuration, we again assume a Dehnen profile but with a potentially different slope $\gamma_f$. The final energy is then
\begin{equation}
E_f=\frac{W_f}{2}
=-\frac{GM^2}{4a_f(5-2\gamma_f)}.
\end{equation}
Assuming conservation of mass and total energy,
$E_i=E_f$, we obtain
\begin{equation}
\frac{a_f}{a_i}
=\frac{1}{2-q_i}
\frac{5-2\gamma_i}{5-2\gamma_f}.
\end{equation}
If the
inner density slope remains unchanged
($\gamma_f=\gamma_i$), the result reduces to
\begin{equation}
a_f=\frac{a_i}{2-q_i}.
\end{equation}
This again reproduces the result we also obtained for the Plummer sphere, showing that it is robust at least under the assumption of the density profile being conserved.

\section{Mergers of star clusters: complete mixing}\label{appendixcomplete}

In Section~\ref{starclustermergers}, we considered the limiting case of incomplete mixing during mergers of star clusters containing normal and massive stars. Here we examine the opposite limit of complete mixing. We assume that the merger remnant is virialized and that both stellar populations follow the same Plummer profile with radius $a_f$. In contrast to the incomplete-mixing approximation, only the \emph{total} energy is conserved; there is no separate energy-conservation equation for each population.

As discussed in Section~\ref{starclustermergers}, the two populations within each progenitor interact gravitationally. Their kinetic energies therefore cannot in general be determined by assigning a self-gravity-based virial parameter of unity to each population separately. Instead, for an initially virialized progenitor $X=A,B$, the total virial theorem gives
\begin{equation}
2(K_{X1}+K_{X2})+W_{X1}+W_{X2}+W_{{\rm int},X}=0,
\end{equation}
where $W_{Xi}=-3\pi GM_{Xi}^2/(32a_{Xi})$ and $W_{{\rm int},X}$ is the mutual gravitational potential energy defined in Eq.~\ref{Wgrav}. Consequently, the total internal energy of progenitor $X$ is
\begin{equation}
E_X=\frac{1}{2}\left(W_{X1}+W_{X2}+W_{{\rm int},X}\right).
\label{eq:complete_progenitor_energy}
\end{equation}
We neglect mass loss and adopt the orbital-energy prescription of Eq.~\ref{Eorbit}. The total mass of the final system is $M_f=M_A+M_B$, where $M_X=M_{X1}+M_{X2}$. Since complete mixing produces a single Plummer density profile, the final virialized energy is
\begin{equation}
E_f=-\frac{3\pi G(M_A+M_B)^2}{64a_f}.
\end{equation}
Energy conservation, $E_A+E_B+E_{\rm orbit}=E_f$, therefore yields
\begin{eqnarray}
&&-\frac{3\pi G}{64}\left(\frac{M_{A1}^2}{a_{A1}}+\frac{M_{A2}^2}{a_{A2}}+\frac{M_{B1}^2}{a_{B1}}+\frac{M_{B2}^2}{a_{B2}}\right)\nonumber\\
&&+\frac{W_{{\rm int},A}+W_{{\rm int},B}}{2}
-f_0\frac{GM_AM_B}{1.305(a_{A1}+a_{B1})}\nonumber\\
&&=-\frac{3\pi G(M_A+M_B)^2}{64a_f}.
\label{eq:complete_energy_balance}
\end{eqnarray}
As in Section~\ref{starclustermergers}, the orbital-energy scale is set by the radii of the normal stellar components. For comparison with Section~\ref{starclustermergers}, we now take $M_{A1}=M_{B1}=M_1$ and $M_{A2}=M_{B2}=M_2$, and set $a_{B1}=f_a a_{A1}$ and $a_{B2}=f_a a_{A2}$. With $a_1=a_{A1}$, $a_2=a_{A2}$, and $M=M_1+M_2$, the gravitational interaction energy scales as $W_{{\rm int},B}=W_{{\rm int},A}/f_a$. Equation~\ref{eq:complete_energy_balance} can thus be written as
\begin{equation}
\frac{1}{a_f}=\frac{1+f_a^{-1}}{4M^2}\left[\frac{M_1^2}{a_1}+\frac{M_2^2}{a_2}-\frac{32W_{{\rm int},A}}{3\pi G}\right]
+\frac{16f_0}{3.915\pi\,a_1(1+f_a)}.
\label{eq:complete_general_radius}
\end{equation}
The interaction term can be evaluated using Eq.~\ref{Wgrav}. In the limit $a_2\ll a_1$, Eq.~\ref{Wint} gives $W_{{\rm int},A}\simeq-GM_1M_2/a_1$, and therefore
\begin{equation}
{\frac{1}{a_f}\simeq
\frac{1+f_a^{-1}}{4(M_1+M_2)^2}\left[\frac{M_1^2}{a_1}+\frac{M_2^2}{a_2}+\frac{32M_1M_2}{3\pi a_1}\right]
+\frac{16f_0}{3.915\pi\,\,a_1(1+f_a)}.}
\label{eq:complete_compact_radius}
\end{equation}
For two identical progenitors ($f_a=1$), this reduces to
\begin{equation}
\frac{1}{a_f}\simeq\frac{1}{2(M_1+M_2)^2}\left[\frac{M_1^2}{a_1}+\frac{M_2^2}{a_2}+\frac{32M_1M_2}{3\pi a_1}\right]
+\frac{8f_0}{3.915\pi\,a_1}.
\label{eq:complete_identical_radius}
\end{equation}
For $M_1=M_2$ and $f_0=0$, this becomes
\begin{equation}
a_f\simeq 8\left[\frac{1}{a_1}+\frac{1}{a_2}+\frac{32}{3\pi a_1}\right]^{-1}.
\end{equation}
If the self-binding energy of the compact massive component dominates, $M_2^2/a_2\gg M_1^2/a_1,\,M_1M_2/a_1$, we obtain $a_f\simeq8a_2$ for $M_1=M_2$. The merger remnant may therefore be much more compact than the original extended component, although complete mixing also redistributes the initially compact massive component over the common final profile.

The limit $M_2\ll M_1$ requires some care: the term $M_2^2/a_2$ is negligible only if the compact component is not so concentrated that its self-binding energy remains important. If $M_2^2/a_2\ll M_1^2/a_1$, Eq.~\ref{eq:complete_compact_radius} gives
\begin{equation}
a_f\simeq\left[\frac{1+f_a^{-1}}{4a_1}+\frac{16f_0}{3.915\pi\,a_1(1+f_a)}\right]^{-1}.
\label{eq:complete_tracer_radius}
\end{equation}
For a parabolic merger ($f_0=0$), this becomes $a_f=4f_a a_1/(1+f_a)$; for identical progenitors ($f_a=1$), it gives $a_f=2a_1$, as required by conservation of the internal binding energies of two initially virialized clusters.

As a further consistency check, if the two populations within each progenitor instead have the same initial radius, $a_2=a_1$, Eq.~\ref{eq:Wint_equal} gives $W_{{\rm int},A}=-3\pi GM_1M_2/(16a_1)$. Substitution into Eq.~\ref{eq:complete_general_radius}, leads to
\begin{equation}
\frac{1}{a_f}=\frac{1+f_a^{-1}}{4a_1}+\frac{16f_0}{3.915\pi\,a_1(1+f_a)},
\end{equation}
independently of the relative masses of the two stellar populations. In particular, $a_f=2a_1$ for $f_a=1$ and $f_0=0$.

\section{Conditions for global violent relaxation in a system of star clusters}
\label{app:global_violent_relaxation}

We consider a stellar system consisting of $N_c$ star clusters,
each with a characteristic mass $M_c$, distributed within a
spherical region of radius $R$. We assume that the cluster
positions are initially random and that their velocities have
random orientations, without any preferred direction or
coherent collective motion. The mean number density of
clusters is defined as
\begin{equation}
n_c = \frac{3N_c}{4\pi R^3}.
\label{eq:vr_cluster_density}
\end{equation}
The corresponding mean mass density is $\rho_c=n_cM_c$. We estimate under which circumstances the motion
of the individual clusters produces sufficiently strong
variations in the global gravitational potential to trigger
violent relaxation. We distinguish such global variations
from local changes in the potential during individual
encounters or mergers, which may still produce local violent
relaxation even if the evolution of the overall system
is more gradual.

At a fixed position $\vec{x}$, the gravitational potential
generated by the clusters, approximated as point masses, is
\begin{equation}
\Phi(\vec{x},t)
=
-\sum_{i=1}^{N_c}
\frac{GM_c}{|\vec{x}-\vec{x}_i(t)|},
\label{eq:vr_total_potential}
\end{equation}
where $\vec{x}_i(t)$ denotes the position of cluster $i$.
Taking the partial time derivative at fixed $\vec{x}$,
we obtain
\begin{equation}
\frac{\partial\Phi}{\partial t}
=
-GM_c\sum_{i=1}^{N_c}
\frac{[\vec{x}-\vec{x}_i(t)]
\cdot\vec{v}_i(t)}
{|\vec{x}-\vec{x}_i(t)|^3},
\label{eq:vr_potential_derivative}
\end{equation}
where $\vec{v}_i$ is the velocity of cluster $i$. We first consider variations in the gravitational potential
on the scale of the entire system, $R$, rather than
fluctuations dominated by close encounters. The
characteristic contribution of an individual cluster
to the large-scale potential is
\begin{equation}
|\Phi_c|\sim\frac{GM_c}{R}.
\label{eq:vr_single_potential}
\end{equation}
The characteristic magnitude of the mean gravitational
potential is then
\begin{equation}
|\overline{\Phi}|
\sim\frac{GN_cM_c}{R}.
\label{eq:vr_mean_potential}
\end{equation}

We define the dynamical time of the entire system as
\begin{equation}
t_{\rm dyn}
\sim
\left(\frac{R^3}{GN_cM_c}\right)^{1/2}
=
\left(\frac{3}{4\pi Gn_cM_c}\right)^{1/2}.
\label{eq:vr_dynamical_time}
\end{equation}
For a system with approximately virial velocities, the
characteristic cluster speed is
\begin{equation}
v_c\sim\left(\frac{GN_cM_c}{R}\right)^{1/2}
\sim\frac{R}{t_{\rm dyn}}.
\label{eq:vr_cluster_velocity}
\end{equation}

\begin{figure}[!h]
    \centering
    \includegraphics{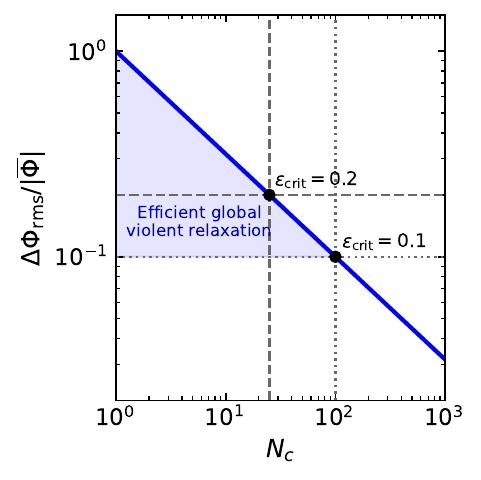}
    \caption{Relative change of the gravitational potential during a dynamical time calculated from Eq.~\ref{eq:vr_relative_potential_change} as a function of the number of star clusters $N_c$, and compared to thresholds of $\epsilon_{\rm crit}=0.1$ and 0.2 for global violent relaxation to happen. The comparison here is only tentative and the threshold parameter should be calibrated via numerical simulations.}
    \label{fig:deltaphi}
\end{figure}

Over one dynamical time, an individual cluster therefore
moves a distance of order $R$. Its contribution to the
large-scale potential can change by an amount of order
\begin{equation}
|\Delta\Phi_c|
\sim\frac{GM_c}{R}.
\label{eq:vr_single_potential_change}
\end{equation}
Because the cluster velocities are initially randomly
oriented, the potential variations produced by different
clusters are not coherent. Assuming that their
contributions are statistically independent, the
root-mean-square variation of the global potential
over one dynamical time is
\begin{equation}
\Delta\Phi_{\rm rms}(t_{\rm dyn})
\sim
\sqrt{N_c}\,|\Delta\Phi_c|
\sim
\frac{GM_c\sqrt{N_c}}{R}.
\label{eq:vr_global_potential_change}
\end{equation}
As we consider only large-scale variations in the
potential, contributions from very close encounters
are excluded from this estimate. Dividing by the mean potential, we obtain
\begin{equation}
\frac{\Delta\Phi_{\rm rms}(t_{\rm dyn})}
{|\overline{\Phi}|}
\sim\frac{1}{\sqrt{N_c}}.
\label{eq:vr_relative_potential_change}
\end{equation}
Thus, the relative variations of the global gravitational
potential decrease as the number of independently moving
clusters increases. Violent relaxation requires sufficiently rapid changes in
the gravitational potential to redistribute orbital
energies over a small number of dynamical times. We
 introduce a dimensionless parameter measuring
the relative variation of the global potential during
one dynamical time,
\begin{equation}
\epsilon_{\Phi}
\equiv
\frac{\Delta\Phi_{\rm rms}(t_{\rm dyn})}
{|\overline{\Phi}|}.
\label{eq:vr_efficiency_parameter}
\end{equation}

We assume that global violent relaxation can become
dynamically significant if the relative potential
variation exceeds a critical value
$\epsilon_{\rm crit}$, such that
\begin{equation}
\epsilon_{\Phi}\gtrsim\epsilon_{\rm crit}.
\label{eq:vr_efficiency_threshold}
\end{equation}
The parameter $\epsilon_{\rm crit}$ represents the
minimum fractional variation of the global gravitational
potential considered sufficient for significant orbital
energy redistribution. Its precise value should likely be calibrated via numerical simulations. Substituting Eq.~(\ref{eq:vr_relative_potential_change})
into Eq.~(\ref{eq:vr_efficiency_threshold}), we obtain
\begin{equation}
\frac{1}{\sqrt{N_c}} \gtrsim \epsilon_{\rm crit}.
\label{eq:vr_number_condition}
\end{equation}
This yields the critical cluster number
\begin{equation}
N_c\lesssim N_{\rm crit}
\equiv\epsilon_{\rm crit}^{-2}.
\label{eq:vr_critical_cluster_number}
\end{equation}

For illustration, a threshold
$\epsilon_{\rm crit}=0.1$ corresponds to
$N_{\rm crit}\sim100$, while
$\epsilon_{\rm crit}=0.2$ gives
$N_{\rm crit}\sim25$. This is illustrated together with Eq.~\ref{eq:vr_relative_potential_change} in Fig.~\ref{fig:deltaphi}. The respective values should be
understood as approximate given the uncertainties on the exact threshold.  Using Eq.~(\ref{eq:vr_cluster_density}), the same
condition can be expressed as
\begin{equation}
n_cR^3
\lesssim
\frac{3}{4\pi\epsilon_{\rm crit}^2}.
\label{eq:vr_density_criterion}
\end{equation}
The characteristic cluster mass $M_c$
cancels from the fractional potential variation
in Eq.~(\ref{eq:vr_relative_potential_change}).

For identical clusters, increasing their masses
strengthens both the mean gravitational potential
and its fluctuations by the same factor. The
cluster mass nevertheless enters the dynamical
time through Eq.~(\ref{eq:vr_dynamical_time}),
and therefore determines the physical timescale
on which the potential variations occur.

The criterion derived here applies to
stochastic large-scale potential variations
generated by initially uncorrelated cluster motions.
It is not a necessary condition for violent
relaxation in general. If the entire system
undergoes coherent gravitational collapse or
another collective instability, the potential
may change by an order-unity fraction within
a dynamical time even for $N_c\gg N_{\rm crit}$.
Conversely, in a statistically stationary
configuration, fluctuations exceeding the
adopted threshold do not by themselves guarantee
efficient violent relaxation.

\end{document}